\documentclass[11pt,a4paper]{article}

\usepackage{amsmath,amsthm}
\usepackage{amssymb}
\usepackage{bm}
\usepackage{mathrsfs}
\usepackage{empheq}
\usepackage{inputenc}
\usepackage{overpic}

\newcommand{\bean}{\begin{eqnarray}}
\newcommand{\eean}{\end{eqnarray}}
\newcommand{\bea}{\begin{eqnarray*}}
\newcommand{\eea}{\end{eqnarray*}}
\newcommand{\beq}{\begin{equation}}
\newcommand{\eeq}{\end{equation}}
\newcommand{\bmat}{\begin{phmatrix}}
\newcommand{\emat}{\end{pmatrix}}

\newcommand{\E}{{\delta \cal E}}
\newcommand{\B}{{\delta \cal B}}
\newcommand{\wq}{q_0}
\newcommand{\tq}{\widehat q}
\newcommand{\tm}{\widehat m}
\newcommand{\wa}{\delta \widetilde A}

\newcommand{\womega}{\widehat \omega}
\newcommand{\da}{{\tau}}

\newcommand{\Q}{{Q_m}}
\newcommand{\Flux}{{Q_m}}
\newcommand{\cQ}{{\cal Q}}
\newcommand{\cM}{{\cal M}}

\newcommand{\bpm}{\begin{pmatrix}}
\newcommand{\epm}{\end{pmatrix}}
\newcommand{\HT}{{T}}
\newcommand{\UnitA}{{\cal A}_l}
\newcommand{\unitA}{{\mathscr A}_0}
\newcommand{\Area}{{\cal A}_r}
\newcommand{\Areah}{{\cal A}_h}
\newcommand{\x}{\zeta}
\newcommand{\Sla}{{a}} 
\newcommand{\Slb}{{b}}
\newcommand{\Slc}{{c}}
\newcommand{\Sld}{{d}}
\newcommand{\tSlb}{b_{\chi_0}}
\newcommand{\tSld}{d_{\chi_0}}
\newcommand{\ia}{{\mathbf a}} 
\newcommand{\ib}{{\mathbf b}}
\newcommand{\ic}{{\mathbf c}}
\newcommand{\id}{{\mathbf d}}
\newcommand{\ca}{{\alpha}} 
\newcommand{\cb}{{\beta}}
\newcommand{\cc}{{\gamma}}

\newcommand{\vtheta}{\vartheta}

\begin{document}

\title{\textbf{Duality and modular symmetry in the quantum Hall effect from Lifshitz holography}}

\author{Brian P. Dolan\footnote{email: bdolan@thphys.nuim.ie} \\ \\
  \textit{Department of Theoretical Physics, Maynooth University}\\
  \textit{Main St., Maynooth, Co.~Kildare, Ireland}\\
  and \\ 
  \textit{School of Theoretical Physics}\\
  \textit{Dublin Institute for Advanced Studies}\\
  \textit{10 Burlington Rd., Dublin, Co. Dublin, Ireland}}
\maketitle

\vspace{-12cm}
\rightline{DIAS-STP-21-05}
\vspace{12cm}

\begin{abstract}
  The temperature dependence of quantum Hall conductivities is studied in the context of the AdS/CMT paradigm using a model with a bulk theory consisting of (3+1)-dimensional Einstein-Maxwell action coupled to a dilaton and an axion, with a negative cosmological constant.
We consider a solution which has a Lifshitz like geometry with a dyonic black-brane in the bulk.
There is an $Sl(2,{\bm R})$ action in the bulk corresponding to electromagnetic duality,
which maps between classical solutions, and is broken to $Sl(2,{\bm Z})$ by Dirac quantisation of dyons.  This bulk $Sl(2,{\bm Z})$ action translates to an action of the modular group on the 2-dimensional transverse conductivities. The temperature dependence of the infra-red conductivities is then linked to modular forms via gradient flow and the resulting flow diagrams show remarkable agreement with existing experimental data on the temperature flow of both integral and fractional quantum Hall conductivities.

\end{abstract}

\newpage

\section{Introduction}

The quantum Hall effect (QHE) is a fascinating phenomenon involving a strongly interacting system that
exhibits an extensive hierarchy of quantum phase transitions. As a strongly interacting quantum system
it is a candidate for testing the ideas of the AdS/CFT correspondence \cite{Maldecana} in the context of a (3+1)-dimensional bulk space-time with a (2+1)-dimensional boundary and there has already been a substantial body of work exploring this possibility \cite{Hartnoll+Kovtun}-\cite{Alejo+Nastase}. 
Although the original AdS/CFT conjecture was for large $N$ supersymmetric theories with 4-dimensional boundary it has been applied to  non-supersymmetric theories in condensed matter (for a review see \cite{HLS}). The QHE is not relativistic nor are the different quantum phases of the system described by conformal field theories (CFT's), but the phase transition between the Hall plateaux are second order and non-relative systems have been considered in the more general framework of geometries with Lifshitz scaling \cite{GKPT} in the AdS/CMT
(Condensed Matter Theory) approach (for a review of Lifshitz holography see \cite{Lifshitz-Holography}).

It was observed in \cite{Shapere+Wilczek}-\cite{Lutken+Ross2} that the QHE has an emergent infra-red modular symmetry relating the different QHE phases and this should be incorporated into any attempt to model the QHE in the AdS/CMT picture.\footnote{The same symmetry was found independently at almost the same time in \cite{KLZ}, though these authors expressed the transformations for the Ohmic and the Hall conductivities separately, complex conductivities were not used.} Some steps have already been taken in this direction \cite{1917,GIKPTW,Fujita:2012fp, Bak+Rey} and these ideas will be explored further here.
In this work gradient flow for the $\beta$-functions for the conductivity will be considered within the AdS/CMT framework. Gradient flow in the QHE compatible with modular symmetry was proposed in \cite{Burgess+Lutken1,Burgess+Lutken2} with potentials that were quasi-holomorphic in the complex conductivity $\sigma=\sigma^{x y}+i \sigma^{x x}$. An anti-holomorphic potential using modular forms was suggested in \cite{L+R:anti-holomorphic1}-\cite{Lutken3} and holomorphic potentials were considered in \cite{Crossover}. For a review of the status of modular symmetry in the QHE see \cite{OLL}.

 The starting point will be a bulk theory with a dilaton and an axion that enjoys $Sl(2,{\bm R})$ symmetry of the
 classical solutions derived from electromagnetic duality \cite{Gibbons+Rasheed}.
 The strength of the coupling between the dilaton $\phi$ and the axion $\chi$ to
 the electromagnetic field is determined by a parameter $\lambda$ and they can be combined into a complex field $\tau=\lambda \chi + i e^{-\lambda \phi}$ which we shall call the dilaxion.
It will be shown that the infra-red conductivity arising from a bulk solution which contains a dyonic black-brane is related
 to the complex conjugate of the value of the dilaxion at the event horizon, $\sigma=-\overline \tau_h$.  Quantum effects then break  $Sl(2,{\bm R})$ to $Sl(2,{\bm Z})$, resulting in rational filling fractions, \cite{Bak+Rey}. The black-brane has a Hawking temperature which allows the temperature dependence of the conductivity 
 to be determined from the classical solution and a flow diagram is generated.

 Assuming anti-holomorphic gradient flow, with a potential that is holomorphic in $\overline\sigma$, and a viable potential function with only one free parameter is constructed using modular  invariants.  There is a second order phase transition between quantum Hall states with a critical exponent that depends on the free parameter in the potential.  The resulting temperature flow of the 2-dimensional conductivity tensor is shown in figure \ref{fig:Gamma02-full-flow}, which compares very favourably with the available
 experimental data \cite{Murzin-integer,Murzin-fractional,Taiwan-group}.

The layout of the paper is as follows.
\S\ref{sec:Lagrangian} describes the background bulk action and dyonic solutions of the equations of
motion are presented in \S\ref{sec:Taylor-solution}.
The Dirac quantisation in the bulk, and how it relates to fractional filling factors on the boundary,
is discussed in \S\ref{sec:DSZ-quantisation}. The RG equation for the conductivity is discussed in \S\ref{sec:Conductivities} where it is shown that the DC conductivity is determined by the value of the dilaxion at
the event horizon. A discussion of the  temperature flow of the DC conductivity is given in \S\ref{sec:T-flow}
and implications of the results are discussed in \S\ref{sec:Discussion}.  Some technical details required
in the main body of the text are given in three appendices.

\section{Einstein-Maxwell-dilaton-axion Lagrangian with a cosmological constant\label{sec:Lagrangian}}

The starting point is Einstein-Maxwell theory in four dimensional space-time with a dilaton, an axion and a negative cosmological constant.  This is interpreted as an effective action action after any charged matter is integrated out and it will suffice for
a preliminary investigation of the resulting 2-point correlators.  The Lagrangian is
\beq {\cal L} = \frac{1 }{2\kappa^2} \bigl({\cal R} -2 \Lambda - \partial^\mu \phi \partial_\mu \phi  -  e^{2\lambda \phi }\partial^\mu \chi \partial_\mu \chi\bigl)
-\frac{1}{4} e^{-\lambda \phi} F_{\mu\nu} F^{\mu\nu} 
+\frac 1 4 \lambda \chi F_{\mu\nu} \widetilde F^{\mu\nu} \label{eq:G-R-lagrangian} \eeq
where  $\kappa^2 = 8 \pi G$, $c=1$, $\lambda$ is a dimensionless parameter,
and \hbox{$\widetilde F^{\mu\nu}= \frac {1}{2}  \frac{\epsilon^{\mu\nu\rho\sigma}}{\sqrt{-g}} F_{\rho\sigma}$}.
In terms of the dilaxion field
$$\da = \lambda\chi + i e^{-\lambda \phi}$$
and 
$$ G^{\mu\nu}  = e^{-\lambda \phi} F^{\mu\nu} - \lambda \chi \widetilde F^{\mu\nu}$$
the Lagrangian is 
\beq {\cal L} = \frac{1 }{2\kappa^2} ({\cal R} -2 \Lambda)  
+ \frac{2}{\kappa^2\lambda^2 }  \frac{\partial^\mu \da \, \partial_\mu \overline \da }{(\da- \overline \da)^2} -\frac 1 4 G^{\mu\nu}F_{\mu\nu}.\label{eq:dilaxion-L}
\eeq
Gibbons and Rasheed \cite{Gibbons+Rasheed} have shown that, given any solution of 
 the equations of motion of (\ref{eq:G-R-lagrangian}), 
new solutions can be generated by applying an $Sl(2,{\bm R})$ symmetry transformation.\footnote{There are no conserved charges associated with this symmetry, it is not  a symmetry of the action.}
Define the complex fields
$$ {\cal F}^{\mu\nu}= F^{\mu\nu}  + i \widetilde F^{\mu\nu},\qquad 
{\cal G}^{\mu\nu}= -i G^{\mu\nu} +  \widetilde G^{\mu\nu}=\overline \tau {\cal F}^{\mu\nu}$$
then an $Sl(2,{\bm R})$ action is defined on the fields by 
$$ \begin{pmatrix}
{\cal G} \cr {\cal F} 
\end{pmatrix} 
\ \longrightarrow \
\bpm {\cal G}' \cr {\cal F}' \epm  = 
\begin{pmatrix}
\Sla & \Slb\cr \Slc& \Sld
\end{pmatrix} 
\begin{pmatrix} {\cal G} \cr {\cal F}
\end{pmatrix}
$$
with
\beq \da \ \longrightarrow \ \da'=\frac{\Sla \da + \Slb}{\Slc\da + \Sld}\label{eq:Sl2R}\eeq
 where $\Sla  $, $\Slb $, $\Slc $ and $\Sld $ are real and  $\Sla \Sld- \Slb\Slc=1$,
while the metric $g_{\mu\nu}$ is left invariant.
If $\bigl(F^{\mu\nu}, \da, g_{\mu\nu}\bigr)$ is a solution then this $Sl(2,{\bm R})$
action generates a new solution $\bigl( (F')^{\mu\nu}, \da', g_{\mu\nu}\bigr)$.

The transformation properties of the electric and magnetic fields can be succinctly written
using an orthonormal basis $\{e^0,e^i\}$, with $i=1,2,3$, 
in which
\[ E_i = -F_{0 i} \qquad \mbox{and} \qquad B^i= \frac{1}{2}\epsilon^{i j k} F_{j k}= -\frac{1}{2}\epsilon^{0 i j k} F_{j k}\]
($\epsilon_{0 1 2 3} = - e^{0 1 2 3}=+1$).
The complex field
\[{\bm E}_i = E_i + i B_i\]
transforms as
\beq {\bm E}_i  \rightarrow {\bm E}'_i = (\Slc \da + \Sld) {\bm E}_i.
\label{eq:Sl2onE}\eeq
The electric charge and total magnetic flux,
\beq
Q_e = \int_{T^2}*G, \qquad \Flux = \int_{T^2} F,\label{eq:*G}\eeq
transform as
\beq \bpm Q_e \\ Q_m \epm \ \rightarrow \ \bpm a & b \\ c & d \\ \epm \bpm Q_e \\ Q_m \epm
\eeq
which is a generalisation of the Witten effect \cite{Witten-effect}.

\section{Static, spherically symmetric solutions \label{sec:Taylor-solution}}

It was shown in \cite{PTW} that there are no static, spherically symmetric solutions
of (\ref{eq:G-R-lagrangian}) when the cosmological constant is positive, but the present
focus will be on the negative case, $\Lambda = -\frac{3}{L^2}$.
A black-brane solution, with a planar
event horizon and negative cosmological constant, was found in \cite{Taylor}. It is a purely electric solution
with the metric\footnote{Equation (3.8) in \cite{Taylor} with some changes in notation.}
\beq
d s^2 = - \left( \frac{r}{l}\right)^{2 z} \left( 1 - \left(\frac{r_h}{r}\right)^{z+2}\right) d  t^2 
+  \left( \frac{l}{r}\right)^2 \frac{d r^2}{ \left( 1 - \left(\frac{r_h}{r}\right)^{z+2}\right)}
 + r^2 (d \vartheta^2 + d \varphi^2) \label{eq:ds2r}
\eeq
where $r_h$, $l$ and $z$ are constants.
This metric gives non-relativistic holography with $z$ a temporal scaling exponent. Such Lifshitz-like geometries were proposed
in \cite{Koroteev+Libanov,KLM} as AdS/CFT models for non-relativistic systems
Constant $r$ hypersurfaces at large $r$ are then conformal to 3-dimensional space-times on which the speed of light is $\left(\frac{r}{l}\right)^{z-1}$, \cite{Ross+Saremi}.
We will use periodic boundary conditions on surfaces of constant $t$ and $r$, in order to keep their area finite, giving them the topology of a torus with area
\beq \Area = r^2 \int_{T^2} d \vartheta \wedge d \varphi.\eeq

The equations of motion for the action (\ref{eq:G-R-lagrangian}) are then satisfied by
\begin{align}
F_{t r}& = -\wq\left(\frac r l \right)^{z+1} \label{eq:F-q}\\
e^{-\lambda\phi} & = e^{-\lambda\phi_0}\left(\frac  l r \right)^4 \label{eq:phi-q}\\
\chi & = \chi_0,\label{eq:chi-q}
\end{align}
with $\wq$, $\phi_0$ and $\chi_0$ constants, provided
\begin{align}
z & = 1+\frac{8}{\lambda^2}\label{eq:z-constraint}\\
l^2 & = \frac{(z+1)(z+2) L^2}{6}\label{eq:u-constraint}\\
\wq^2 e^{-\lambda\phi_0}& = \frac{6}{\kappa^2 L^2}\frac{(z-1)}{(z+1)}.
\end{align}
$z$ and $l$ are 
not free, they are fixed in term of $\lambda$ and $L$, and $q_0$ and $\phi_0$ are not independent.
The Lifshitz scaling symmetry is broken by the $r$-dependence of $\phi$, \cite{Taylor-Lifshitz}.
For large $r$ the Ricci tensor is
\[R_{\mu\nu} \ \rightarrow \ \frac {1} {l^2}\bpm  z(z+2) \left(\frac{r}{l} \right)^{2 z} & 0 & 0 & 0 \\
0 & -(z^2+2) \left(\frac l r  \right)^2 & 0 & 0 \\
0 & 0 & -(z+2) r^2 & 0 \\
0 & 0 & 0 & -(z+2)  r^2 \\\epm  \] 
and the metric is only asymptotically AdS if $z=1$.

Although $F_{t r}$ and $\phi$ diverge asymptotically
\[ G_{t r}=-\wq e^{-\lambda\phi_0}\left(\frac r l \right)^{z-3}, \qquad  G^{t r}=\wq e^{-\lambda\phi_0}\left(\frac r l \right)^{-(z+1)},  \]
the energy density 
\[ -G^{\mu\nu} F_{\mu \nu} =  2 \wq^2 e^{-\lambda\phi_0}\]
is finite. The pre-factor $e^{-\lambda \phi(r)}$ is like a background electric permittivity,\footnote{At the same time it is  an inverse background magnetic susceptibility --- it does not affect the speed of light.} which dies off
asymptotically so as to render the energy density in the electric field finite at large $r$,
although the electric field itself diverges there. 

The total electric charge on the torus can be calculated using Gauss' law
 and is independent of $r$,
\beq Q_e = - \wq  e^{-\lambda \phi_0} l^2
\int_{T^2} d\vartheta \wedge d\varphi\label{eq:Qe-r}\eeq
(the normal to the torus is taken to be in the direction of decreasing $r$).
There is no magnetic charge and this will be called the electric solution.

\subsection{Hawking temperature \label{sec:Hawking-Temperature}}

The Hawking temperature associate with the metric (\ref{eq:ds2r}) is
\beq \HT=\frac{(z+2)\hbar}{4 \pi l} \left(\frac{r_h}{l}\right)^z.\label{eq:T_H_l}\eeq
Thus $\HT\rightarrow  0$ smoothly as $r_h\rightarrow 0$ and there is no Hawking-Page phase transition associated with this geometry. 

The solution fixes $l$ in terms of the Lagrangian parameters $\lambda$ and $L$,
but $r_h$ is a free parameter which is related to the mass of the black-brane.
The area of the event horizon is
\beq \Areah = \frac{r_h^2}{l^2}  \UnitA,\eeq
where
\beq\UnitA=l^2 \int_{T^2}d\vartheta\wedge d\varphi\label{eq:UnitA-def}\eeq is a fixed fiducial area,
in terms of which the Bekenstein-Hawking  entropy is
\[ S=\frac{\Areah}{4 G \hbar} = \frac{2 \pi  \UnitA  r_h^2}{\kappa^2 \hbar\, l^2}
= \frac{2 \pi \UnitA}{\kappa^2 \hbar} 
\left(\frac{4 \pi l \HT}{(z+2)\hbar}\right)^{2/z}.\]
From the first law of black hole thermodynamics
\[
d M  = T d S =  T \left(\frac{4 \pi  \UnitA  r_h}{\kappa^2 \hbar\,l^2} d r_h\right)
  = \frac{(z+2) r_h^{z+1} \UnitA}{l^{z+3} \kappa^2}d r_h =
  d \left(\frac{ \UnitA r_h^{z+2}}{l^{z+3} \kappa^2}\right),\] 
so
\beq M=\frac{ \UnitA}{\kappa^2 l}
\left(\frac{r_h}{l} \right)^{z+2}\label{eq:M}.\eeq
The heat capacity is 
\[ C_p = T \frac{\partial M}{\partial T} 
= \frac{(z+2)}{z} M 
= \frac{(z+2)}{z}  \frac{ \UnitA}{\kappa^2 l}
\left(\frac{4 \pi l T}{(z+2)\hbar } \right)^{(z+2)/z}
> 0 \]
and the system is thermodynamically stable for any $T>0$, in agreement with the observation above that there is no Hawking-Page phase transition.

For the classical solution to be valid we must ensure that both $l$ and $r_h$ are well
above the Planck length $L^2_{Pl}=\frac{\hbar \kappa^2}{8 \pi}$,
\[ \frac{l^2} {\kappa^2}  \gg \hbar,\qquad \frac{r_h ^2} {\kappa^2} \gg \hbar,\]
but this in itself does not restrict $\frac{r_h}{l}$.  Nevertheless
\[ T = \frac{(z+2)}{4 \pi } \left(\frac{L_{Pl}}{l}\right) \left( \frac{r_h}{l} \right)^z  T_{Pl},  \]
where $T_{Pl}=\frac{\hbar}{L_{Pl}}$ is the Plank temperature  so demanding that $T \ll T_{Pl}$ imposes the extra condition
\beq
\left( \frac{r_h}{l} \right)^z \ll \frac{l}{L_{Pl}}.
\label{eq:r_h-over-l-constraint}\eeq

\subsection{Dyonic solutions\label{sec:Dyon}}

The $Sl(2,\bm R)$ action can now be used to generate static, 
spherically symmetric dyonic solutions from (\ref{eq:ds2r})-(\ref{eq:chi-q}).
It is convenient to first change co-ordinates: let
$\tilde t= \bigl(\frac{r_h}{l}\bigr)^z t$, $u=\frac{r_h}{r}$,
$x=r_h d\vartheta$, and $y=r_h d\varphi$, 
so $0\le u \le 1$ and event horizon is at $u=1$, the asymptotic region is $u\rightarrow 0$.
Then (\ref{eq:ds2r}) becomes
\beq
d s^2 =  -\left( 1 - u^{z+2}\right) \frac{d \tilde t\,^2}{{u^{2z}}} 
 + \frac{l^2 }{(1-u^{z+2})}\frac{d u^2 }{u^2} + \frac{d x^2}{u^2} + \frac{d y^2}{u^2}.
\label{eq:ds2u}
\eeq
In these variables the electric solution (\ref{eq:F-q}), (\ref{eq:phi-q}) and (\ref{eq:chi-q}) is
\begin{align}
  F_{\tilde t u}&=\wq\left(\frac{r_h}{l}\right)^2 \frac{l}{u^{z+3}} \label{eq:F-q-u}\\
  G_{\tilde t u}&=\wq e^{-\lambda \phi_0}  \left( \frac{l}{r_h} \right)^2 \frac{l}{u^{z-1}},           
           \label{eq:G-q-u}  \\
  e^{-\lambda\phi}  & = e^{-\lambda \phi_0}\left( \frac{l}{r_h} \right)^4 u^4. \label{eq:phi-q-u}
\end{align}
In terms of orthonormal 1-forms
\[
  e^0=\frac{\sqrt{1-u^{z+2}}}{u^z} d \tilde t, \quad
  e^1 = \frac{l}{u\sqrt{1-u^{z+2}}} d u, \quad
  e^2 = \frac{d x}{u}, \quad   e^3 = \frac{d y}{u},
\]
this purely electric configuration is
\beq F =  \left(\frac{r_h}{l}\right)^2\frac{\wq}{u^2} e^{0 1}, \qquad
  G =  \wq\left( \frac{r_h}{l} \right)^2
   \x u^2 e^{0 1},
  \qquad \da= \lambda \chi_0 +i   \x u^4 \label{eq:electric-sol-u}\eeq
  where $e^{0 1} = e^0 \wedge e^1$ and
  \beq \x=e^{-\lambda \phi_0} \left( \frac{l}{r_h} \right)^4.\label{x-def}\eeq

Under a $Sl(2,{\bm R})$ transformation $\bpm  \Sla &  \Slb\\  \Slc&  \Sld\epm$ 
the metric is unchanged and, from (\ref{eq:Sl2onE}), the purely electric configuration (\ref{eq:electric-sol-u}) is mapped to a configuration with a constant magnetic component,
\begin{align}
  F & = \wq\left( \frac{r_h}{l} \right)^2 \left(\frac{\Sld}{u^2} e^{0 1} - \Slc \, \x u^2 e^{2 3} \right)\label{eq:F-dyon}\\
  G & =\wq\left( \frac{r_h}{l} \right)^2 \left(\Sla\,  \x u^2 e^{0 1} +
      \frac{\Slb} {u^2} e^{2 3} \right),
  \end{align} 
  while the dilaxion becomes
\beq
\da = \frac{i \Sla  \x u^4 + \Slb + \Sla \lambda \chi_0}
{i\Slc  \x u^4\ + \Sld + \Slc\lambda\chi_0}
= \frac{i \Sla  \x u^4 + \tSlb} {i\Slc  \x u^4\ + \tSld }\label{eq:dyonic-da}
\eeq
where $\tSlb = b + \Sla \lambda \chi_0$ and $\tSld = \Sld + c \lambda \chi_0$
(with $\Sla \tSld - \tSlb \Slc =1$).
The dilaton and axion fields are separately 
\beq e^{-\lambda\phi}  =  \frac{ \x u^4} { \tSld^2 +  \Slc^2  \x^2 u^8  },
\qquad \lambda \chi  =
\frac{ \tSlb\tSld+ \Sla\, \Slc\,  \x^2 u^8}
{\tSld^2 + \Slc^2  \x^2 u^8}.\label{eq:phi-chi}
\eeq
Defining $q=q_0 d$ and  $m = -\wq\,  c\, e^{-\lambda \phi_0}$ the Maxwell field strength is   
\beq F =\left(\frac{r_h}{l}\right)^2 \frac{q}{u^2}e^{0 1} +
\left(\frac{l}{r_h}\right)^2 m u^2 e^{2 3},\label{eq:F-u}
\eeq
and the full set of coupled equations of motion with
(\ref{eq:F-u}) and (\ref{eq:phi-chi})
are satisfied in the metric (\ref{eq:ds2u}) provided 
\begin{align}
\kappa^2 l^2 e^{-\lambda \phi_0} q^2 &= (z-1)(z+2)\, \Sld^2,  \label{eq:qm-delta-gamma-I}\\
\kappa^2 l^2 e^{\lambda \phi_0} m^2 &= (z-1)(z+2)\, \Slc^2,\label{eq:qm-delta-gamma-II}\\
e^{-\lambda \phi_0} q \Slc + m \Sld&=0. \label{eq:qm-delta-gamma-III}
\end{align}

If $\phi_0=0$ in the electric solution a non-zero $\phi_0$ can be
generated by choosing
$b=c=0$ and $a = d^{-1} = e^{\lambda \phi_0/2}$, and if $\chi_0=0$ a constant non-zero topological susceptibility is generated by choosing $c=0$ and $b\ne 0$.
Apart from $r_h$ there are four parameters in the solution: physically these are $q$, $\phi_0$, $m$
and $\chi_0$, equivalent to $a$, $b$, $c$ and $d$, but only three of these are independent 
since $a d - b c=1$.

  The total magnetic flux of the solution is
\beq \Flux = \int_{T^2} F =   m \left(\frac{l}{r_h}\right)^2 \Areah=m l^2 \unitA \label{eq:Q_m}\eeq
where $\unitA=\int_{T^2} d \vartheta  \wedge d \varphi$ is dimensionless.
The electric charge is
\beq
Q_e =  \int_{T^2} *G 
= -e^{-\lambda \phi_0} \left(\frac{l}{r_h}\right)^2 \frac{q  \Sla }{\Sld} \Areah= 
\frac{ m \Sla  }{\Slc } \UnitA,
\label{eq:Q_e}
\eeq
where the second equality follows from (\ref{eq:qm-delta-gamma-III}).

If the total charges $Q_m$ and $Q_e$ are kept constant the magnetic field and charge density are functions of $u$
\[B(u) = \frac{Q_m}{\Areah} u^2
  \qquad \mbox{and} \qquad
  \rho(u)= \frac{Q_e}{\Areah} u^2.\]
At the event horizon
\beq B_h= \frac{Q_m}{\UnitA} \frac{l^2}{r_h^2} =\frac{Q_m}{\UnitA}\left(\frac{(z+2)\hbar}{4 \pi l T}\right)^{2/z}\label{eq:B-T}\eeq
and
\beq \rho_h= \frac{Q_e}{\UnitA} \frac{l^2}{r_h^2} =\frac{Q_e}{\UnitA}\left(\frac{(z+2)\hbar}{4 \pi l T}\right)^{2/z}.\label{eq:rho-T}\eeq

In the quantum Hall effect, with electric charge $e_0$ and unit of magnetic flux $\Phi_0$,
the filling factor is defined to be
\beq \nu
= \frac{\rho}{e_0}\frac{\Phi_0}{B}
=\frac{Q_e}{\Q} \frac{\Phi_0}{e_0}
=\frac{\Sla}{\Slc}\frac{\Phi_0}{e_0} \label{eq:filling-factor}\eeq
and is independent of $u$.

\section{Dirac-Schwinger-Zwanziger quantisation \label{sec:DSZ-quantisation}}

It was observed in \cite{Bak+Rey} that the Dirac-Schwinger-Zwanziger quantisation condition on dyons in the bulk translates to a rational filling factor on the boundary.
In the AdS/CFT correspondence gauge symmetries in the bulk correspond to global symmetries
on the boundary and the authors of \cite{Bak+Rey} identify the global $U(1)$ on the boundary
arising from the bulk $U(1)$ gauge field as being associated with the conservation of composite fermion number in the QHE on the boundary. In Jain's composite fermion picture of the QHE \cite{Jain-I}, \cite{Jain-II} the statistical
gauge field generates a \lq fictitious' background magnetic field which is quantised and concentrated in $\delta$-function magnetic magnetic vortices.

For the general dyonic solution  (\ref{eq:F-u})-(\ref{eq:qm-delta-gamma-III})  the total magnetic flux through the torus $T^2$,
at fixed $t$ and $u$, is (\ref{eq:Q_m}).
  If there is a quantum unit of magnetic flux $\Phi_0$ then $\Flux$ will be integer multiple of $\Phi_0$
\[ \Flux = N_m  \Phi_0, \qquad N_m  \in {\bf Z}.\]
Mathematically one could set $\Phi_0=2 \pi$ so that
\[ N_m = \frac{1}{2\pi} \int_{T^2} F \]
is the first Chern number of a $U(1)$ line bundle over the torus.
In physical units $\Phi_0= \frac{2 \pi \hbar}{e}$, where $e$ is the charge of the electron,
might seem natural and indeed this would be correct for a quantum Hall system, but there are other possibilities.  In a superconductor, for example,  Cooper pairs have charge $e_0=2e$
and the unit of magnetic flux is $\frac {h}{ 2 e}$. Of course $\frac{\Phi_0}{e_0}$ is dimensionless
and in the following we shall use units with $\frac{e_0^2}{h}=\frac{n^2 e^2}{h}=1$, where $n=1$ for quantum Hall systems and $n=2$ for superconductors.
The unit of electric charge is then $e_0=n e$ and the unit of magnetic flux  
is $\Phi_0=\frac{h}{n e}$, with $e_0 \Phi_0=h$ and $\frac{e_0}{\Phi_0}=1$,
in these units $e_0=\Phi_0=\sqrt{2 \pi \hbar}$.

The total electric charge $Q_e$ is a multiple of $e_0$
\[ Q_e =N_e e_0, \qquad N_e \in {\bf Z},\]
and, from the Dirac quantisation condition,
\[ \frac{Q_e \Q}{2 \pi \hbar} =N_e N_m  \in {\bf Z}.\]

Eliminating $q $ and $m$ in favour of $\Sld $ and $\Slc $ using 
(\ref{eq:qm-delta-gamma-I})-(\ref{eq:qm-delta-gamma-III}), equations (\ref{eq:Q_m}) and (\ref{eq:Q_e}) give
\beq
Q_m  = -\frac{\Slc \,e^{-\lambda\phi_0/2}}{\eta} ,\qquad Q_e  = -\frac{\Sla\, e^{-\lambda \phi_0/2} }{\eta},
\label{eq:Q_m+Q_e}\eeq
where
\beq \eta  = \frac{\kappa l }{\UnitA}\frac{1}{\sqrt{(z-1)(z+2)}}.
\label{eq:eta-def}\eeq
Let $N$ be the greatest common divisor of $N_e$ and $N_m$,
where the sign chosen so that $N_e= -N \ia$ and $N_m=-N \ic$,
with $\ia$ and $\ic$ mutually prime. Then
\[
Q_m  = -N  \ic \Phi_0  ,
\qquad Q_e  = -N \ia \,  e_0 
\]
with
\[\ia = \frac{\Sla e^{-\lambda \phi_0/2} }{\sqrt{2 \pi \hbar}}\left(\frac{1}{\eta N}\right),
  \qquad \ic= \frac{\Slc e^{-\lambda \phi_0/2} }{\sqrt{2 \pi \hbar}}\left(\frac{1}{\eta N}\right).\]

If we further define
\beq
\ib = \Slb  e^{\lambda \phi_0/2}  \eta N\sqrt{2\pi\hbar},\qquad
\id =  \Sld    e^{\lambda \phi_0/2}\eta N \sqrt{2 \pi \hbar}
\eeq
then 
\beq
\begin{pmatrix} \Sla & \Slb\\ \Slc& \Sld
\end{pmatrix}=
\begin{pmatrix} \ia & \ib \\ 
  \ic  &  \id \end{pmatrix}
\begin{pmatrix}  e^{\lambda \phi_0/2} \eta N \sqrt{2\pi \hbar } & 0 \\ 0 &  \bigl(e^{\lambda \phi_0/2}  \eta N \sqrt{2\pi \hbar } \bigr)^{-1}  \end{pmatrix}
\label{eq:DSZ-quantisation}\eeq
with $\ia  \id - \ib \ic=1$.
Setting $\bpm \ia & \ib \\ \ic & \id \epm = \bpm 1 & 0 \\ 0 & 1 \epm$
in (\ref{eq:DSZ-quantisation}) equation (\ref{eq:dyonic-da})
then gives the electric dilaxion field to be
\[\tau =2 \pi\hbar N^2 \eta^2 e^{\lambda \phi_0} ( \lambda \chi_0 +    i \x u^4 ),\]
so we  set
\beq e^{-\lambda \phi_0}=2 \pi \hbar  N^2 \eta^2
= \frac{2 \pi \hbar \kappa^2 }{(z-1)(z+2)l^2} \left(\frac{N}{\unitA}\right)^2,\label{eq:phi0-N}\eeq
this is does not reduce the number of parameters in the solution, we are just trading $\phi_0$ for $\frac{N}{\unitA}$.
The dilaxion associated with any magnetic monopole or dyon solution compatible
with the Dirac quantisation condition is now obtained from
\[\da = \lambda \chi_0 + i \zeta u^4 \] by acting on it by a general element of $\Gamma(1)$.

Note that an immediate consequence of Dirac-Schwinger-Zwanziger quantisation 
is that the filling fraction
\beq \nu = \frac{\Sla}{\Slc} = \frac{\ia}{\ic}\label{eq:rational-nu}\eeq
is a rational number.

In summary the general quantised dyonic solution in terms of the integers $\ia$, $\ib$, $\ic$ and $\id$
satisfying $\ia  \id - \ib \ic=1$, is
\begin{align}
Q_e&=-N\ia\,  e_0, \qquad \qquad\quad   \Q = -N \ic\Phi_0, \label{eq:sol-I} \\
  F_{\tilde t u}  &= \left(\frac{r_h}{l}\right)^2\frac{q \,l}{u^{z+3}},\qquad 
                           \ F_{ x y}=  m  \left(\frac{l}{r_h}\right)^2, \label{eq:sol-II}\\
                    q  &=\id \left(\frac{ (z-1)(z+2) }{2 \pi \hbar \kappa^2 l^2 }\frac{\UnitA} {N}\right) e_0 , 
\qquad m=  -\ic \left(\frac{N  \Phi_0}{\UnitA}\right)  ,\label{eq:sol-III}\\
  \da & =\left(\frac{ \ia( \lambda \chi_0  + i \x  u^4)  + \ib }
        {\ic (\lambda \chi_0 + i \x u^4) + \id }\right) \frac{e_0}{\Phi_0},\label{eq:sol-IV}
\end{align}
with $\x = 2\pi\hbar N^2\eta^2 \left(\frac{l}{r_h} \right)^4$ and $e_0=\Phi_0=\sqrt{2 \pi \hbar}$.

The full solution is acted on by $Sl(2,{\bm Z})$ but,  apart from the sign of $F_{\mu\nu}$, the dyon solution is invariant under 
\[
\begingroup
\renewcommand*{\arraystretch}{1.3}
  \begin{pmatrix}
    \ia  & \ib  \\
  \ic   & \id
\end{pmatrix}
\rightarrow
- \begin{pmatrix}
  \ia  & \ib   \\
\ic   & \id
\end{pmatrix},
\endgroup
\]
and the dilaxion transforms under the modular group $\Gamma(1) \approx Sl(2,{\bf Z})/{\bf Z}_2$.
Note also that, while $m$ is proportional to $\ic $ and $q$ is proportional to $\id $,
 the electric charge is determined by $\ia $ and not by $\id $, because of the Witten effect.

The dimensionless parameter $\x$ depends on the Hawking temperature of the 
system (\ref{eq:T_H_l}).
Since 
\[  \frac{r_h}{l} =  
\left\{\frac{4 \pi l T}{ (z+2) \hbar} \right\}^{1/z}\]
we can write
\[\x =\left(\frac{T_*}{T}\right)^{4/z}=\Theta^{-4/z} \quad \mbox{with} \quad \Theta = \frac{T}{T_*}\] 
where
\begin{align*}
  T_* &= \left(\frac{(z+2)\hbar}{4 \pi l}\right) \left(2 \pi \hbar N^2  \eta^2 \right)^{z/4}\\
  &=\left(\frac{z+2}{4\pi}\right) \left(\frac{N^2}{4\unitA^2(z-1)(z+2)}\right)^{z/4}
  \left( \frac{L_{Pl}}{l}\right)^{(z+2)/2} T_{Pl}.\end{align*}
In the next section $T_*$ will be related to the critical temperature of
a second order phase transition and, if it is to be interpreted as a physical temperature, we should therefore ensure that $T_* \ll T_{Pl}$ which requires
\[\frac{N^2}{\unitA^2} \ll
  \left( \frac{l}{L_{Pl}}\right)^{2(z+2)/z}.  \] 
At the same time $T\ll T_{Pl}$ implies that
\[ \Theta^{4/z}  \ll \frac{\unitA^2}{N^2}  \left( \frac{l}{L_{Pl}}\right)^{2(z+2)/z}. \]

\section{Conductivities\label{sec:Conductivities}}

In the AdS/CMT paradigm the boundary of space-time is associated with a $2+1$ dimensional
system which we shall interpret in the present context as a strongly coupled electron system. This is perhaps rather radical for the classical dyon solution solution presented in the previous
section, as the bulk space-time is not even asymptotically AdS, but we shall see that interesting results emerge notwithstanding (our model can be viewed as the near horizon limit of an asymptotically AdS model \cite{GIKPTW}).
The presence of temporal scaling, with $z>1$,
results in a boundary theory which is non-relativistic and the non-zero magnetic
 field generated by the dyon sets the stage for the analysis of the quantum Hall effect.

The conductivity tensor is obtained in linear response from a variation
 \beq\delta E_\ca(\tilde t,u)=e^{-i\widetilde \omega \tilde t} \delta \widetilde E_\ca(u),
   \qquad \delta B(\tilde t,u)^\ca= e^{-i\widetilde \omega \tilde t} \delta \widetilde B^\ca(u)\label{eq:deltaE deltaB}\eeq
   in the transverse electric and magnetic fields ($\ca=x,y$) which satisfies the linearised equations of motion in the chosen background.  We work in  a gauge in which the potential for the field variation is
  \[\delta A_\ca(\tilde t,u) = e^{-i\widetilde \omega \tilde t}\wa_\ca(u), \]
  where it is understood that $\wa_\ca(u)$ depends on $\widetilde \omega$.
  A static electric field can be modelled either with
   \beq  \delta A_\ca(\tilde t,u) = -(\delta E_\ca^0) \tilde t + \delta \widetilde A_\ca(u),\label{eq:DC-A}\eeq
   provided $\delta E_\ca^0$ is independent of $u$, or by allowing $\wa_\ca$ to have a pole at $\widetilde \omega=0$
   with residue $-i \delta E_\ca^0$.

 The electric field generates a current
   \[ \delta J^\ca = \sigma^{\ca \cb} \delta E_\ca
     \qquad \Rightarrow \qquad \sigma^{\ca \cb}
     = \frac{\delta J^\ca}{\delta E_\cb}\]
  where
   \[ \delta E_\ca = - \delta F_{\tilde t \ca}=i \widetilde \omega \wa_\ca,
     \qquad
     \delta B^\ca = \frac{\epsilon^{\tilde t u \ca  \cb}}{\sqrt{-g}}\delta  F_{u \cb}
     = -\frac{1}{l}u^{z+3} \epsilon^{\ca \cb}\wa'_\cb.\]
   The current at any given $u$ is obtained from the action by varying $\wa_\alpha(u)$ keeping the potential
   at the event horizon fixed,
   \[ \widetilde J^\ca(u) = \frac{\delta S[A]}{\delta \widetilde A^\ca(u)} \]
   from which the transverse conductivity tensor is \cite{H-H}
   \beq \sigma^{\ca \cb}(\widetilde \omega,u) = \frac{1}{i\widetilde \omega}\frac{\delta^2 S[A]}{\delta \widetilde A^\ca(u) \delta \widetilde A^\cb(u)},\label{eq:Kubo}\eeq
   
   If we wish to take quantum corrections into account the classical action $S[A]$ should be replaced by an effective
   quantum action $S_{eff}[A]$, but the philosophy of the AdS/CFT correspondence is that a classical solution in the bulk corresponds to a strongly
   interacting quantum theory on the boundary so
   this conductivity will be interpreted as the conductivity of a strongly interacting quantum system at scale $r=\frac{r_h}{u}$. Bulk quantum corrections would be a further refinement.

It will be convenient to use complex co-ordinates and define
\begin{align}
  \sigma_\pm & = \sigma^{x y} \pm i  \sigma^{x x} \label{eq:sigma-pm}\\
  \delta J_\pm & =\delta J^x \pm i \delta J^y\\
\E_\pm & = \delta E_x \pm i \delta E_y \label {eq:delta-Epm-def}\\
\B_\pm &= \B^x \pm i \B^y=\frac{(1-u^{z+2})}{u^{2(z+1)}}(\delta B^x \pm i \delta B^y).\label{eq:delta-Bs-def}
\end{align}
The pre-factor in (\ref{eq:delta-Bs-def}) ensures that, in an orthonormal basis,
where the electric and magnetic fields have components $\delta E_i$ and $\delta B^i$ with $i=2,3$,
\beq \E_\pm = \frac{\sqrt{1-u^{z+2}}}{u^{z+1}}(\delta E_2 \pm i \delta E_3),
  \quad \B_\pm = \frac{\sqrt{1-u^{z+2}}}{u^{z+1}}(\delta B^2 \pm i \delta B^3),\label{eq:delta-Es-def}\eeq
  so $\E_\pm$ and $\B_\pm$ have the same weight.

Adapting the analysis in \cite{H-H} to include the dilaxion
it is shown in appendix \ref{app:LRT} that the conductivity is\footnote{The pre-factors in (\ref{eq:delta-Es-def}) cancel in this definition and $\sigma_\pm$ is a ratio of two quantities evaluated in an orthonormal basis, unlike (\ref{eq:Kubo}) this definition is independent of how the time co-ordinate is scaled, it does not matter whether $t$ or $\tilde t$ is used.}
\begin{equation}
  \sigma_\pm(u) = e^{-\lambda \phi} \frac{\B_\pm}{\E_\pm}-\lambda \chi,
  \label{eq:sigma(u)}
\end{equation}
in units with $\frac{e_0^2}{h}=1$.

\subsection{DC conductivity\label{sec:DC-conductivity}}

When $\widetilde \omega=0$  the DC conductivity (\ref{eq:sigma(u)}) can be obtained
directly from  Maxwell's equations for $\delta F_{\mu\nu}$ in the dyon background,
\beq
\partial_\mu\bigl(\sqrt{-g}\, e^{-\lambda\phi} \delta F^{\mu\nu}\bigr)-\frac{\lambda}{2} \epsilon^{\mu\nu\sigma\lambda} (\partial_\mu \chi) \delta F_{\sigma\lambda}=0.\label{eq:full-Maxwell}
\eeq
For static, homogeneous variations all fields are independent of time and independent of $x$, and $y$
so, we only need consider transverse co-ordinates $\nu=\ca=x$ or $y$ for which
\[\bigl(\sqrt{-g}\, e^{-\lambda\phi} \delta F^{u \ca}\bigr)' 
  + \lambda \epsilon^{\ca \cb} \chi' \delta F_{\tilde t\cb} =0\]
where $'=\frac{d}{d u}$.
For the DC conductivity
\[ \delta F_{\tilde t \cb} = - \delta E^0_\cb\]
are constant and Maxwell's equations reduce to 
\beq\bigl( e^{-\lambda \phi}\B_\pm - \lambda \chi \E^0_\pm\bigr)'=0.\label{eq:DC-Maxwell}  \eeq

We use ingoing boundary conditions at
the event horizon (this is the source of dissipation in the AdS/CFT paradigm \cite{KSS}).
With Eddington-Finkelstein co-ordinates
\[ v_\pm = \tilde t \pm \int \sqrt{\frac{g_{u u}}{-g_{\tilde t \tilde t}}} \,d u =
  \tilde t \pm l \int \frac{ u^{z-1} d u}{(1-u^{z+2})} \]
ingoing boundary conditions at $u=1$ require that
\beq
  \left.\partial_{v_+} (\delta A_\pm)\right|_{u=1} =0\eeq
  so, near $u=1$, 
\beq \mp i u^{z-1} l \B_\pm  = (1-u^{z+2}) \delta A_\pm' \  \approx - l \partial_{\tilde t} (\delta A_\pm) = l (\E^0_\pm) \label{eq:ingoing-BC}
\eeq
where $\delta  A_\pm = \delta  A_x \pm i \delta  A_y$.

Integrating Maxwell's equations (\ref{eq:DC-Maxwell})
\begin{align}
  \frac{e^{-\lambda \phi}\bigl(1-u^{z+2}\bigr) }{l u^{z-1}}(\delta  A_\pm)'
  \pm i\lambda\chi (\E^0_\pm) &= C_\pm\\
  \Leftrightarrow \qquad  e^{-\lambda \phi}(\B_\pm)  - \lambda\chi (\E^0_\pm) &= \pm i C_\pm
  \label{eq:integrated-Maxwell}
\end{align}
with $C_\pm = C^x \pm i C^y$ constants.
Hence the DC conductivity
\[ \sigma_\pm=e^{-\lambda \phi}\frac{\B_\pm}{\E^0_\pm} - \lambda \chi = \pm i\frac{C_\pm}{ \E^0_\pm}. \]
is independent of $u$.
The constants $C_\pm$ can be obtained from Maxwell's equations in the dyon background,
\beq
  \x (1-u^{z+2})\delta \widetilde A_\pm'  =
\bigl\{C_\pm(\tSld^2 + \Slc^2  \x^2 u^8)
  \mp i (\tSlb \tSld + \Sla \Slc  \x^2 u^8)  (\E^0_\pm)\bigr\} l u^{z-5}  
  \label{eq:delta-A'}\eeq
  and the ingoing boundary condition (\ref{eq:ingoing-BC}) enforces
  \begin{align*}
    C_\pm & = \pm i
            \left(\frac{\tSlb \tSld + \Sla \Slc  \x^2 }{\tSld^2 + \Slc^2  \x^2 }\right)\E^0_\pm
            + \left(\frac{ \x}{\tSld^2 + \Slc^2  \x^2} \right)\E^0_\pm\\
\Rightarrow \qquad C_+ & = C_-^* = i \,\overline \da_h\E^0_+.\end{align*}
where $\da_h$ is the dilaxion at the event horizon.
Hence the DC conductivities are given by the value of the dilaxion field at the horizon,
\beq\sigma_+ = - \overline \da_h, \qquad  \sigma_-= - \da_h.\label{eq:sigma-DC}\eeq

Maxwell's equations can also be solved to give  $\B_\pm(u)$ explicitly in the static case, 
\begin{align}
  (1-u^{z+2})\delta \widetilde A_\pm' & =
  \frac{ (\tSld \pm i  \x \Slc)(\tSld \mp i\Slc  \x u^8)}{(\tSld^2 + \x^2\Slc^2)} \, u^{z-5}l (\E^0_\pm)\nonumber\\
\Rightarrow \qquad  \B_\pm & = \pm i\,
  \frac{ (\tSld \pm i  \x \Slc)(\tSld \mp i\Slc  \x u^8)}{u^4(\tSld^2 + \Slc^2  \x^2)}
  (\E^0_\pm).\label{eq:DC-B}\end{align}

Note that, although  equation (\ref{eq:DC-B}) implies that
$\B_\pm$ diverges as $1/u^4$ as $u \rightarrow 0$, in an orthonormal
basis (\ref{eq:delta-Es-def})
\[ \delta B^i \sim u^{z-3}, \]
which is finite for $z\ge 3$. Demanding that the energy density
in the magnetic field perturbation
\[ \frac{e^{-\lambda \phi}}{2} (\B_i)^2 \approx u^4 u^{2(z-3)} \]
is finite as $u\rightarrow 0$ in a local inertial reference frame gives the weaker
condition $z\ge 1$.

 \subsection{AC conductivity}

 For oscillating perturbations
Maxwell's equations (\ref{eq:full-Maxwell}) give us
\beq
\pm\frac{\womega u^{z-1} }{(1-u^{z+2})}(\E_\pm) +(e^{-\lambda \phi} \B_\pm)' -\lambda \chi' (\E_\pm) =0,
\eeq
where $\womega = l \widetilde \omega$.

Consequently the conductivity is no longer independent of $u$, but instead
\[ \sigma_\pm' = - e^{-\lambda \phi}  \left(\frac{\B_\pm  \E_\pm'}{\E_\pm^2}\right)  \mp\frac{ \womega u^{z-1} }{(1-u^{z+2})}
  = - \left(\sigma_\pm + \lambda \chi\right) \left(\frac{\E_\pm'}{\E_\pm}\right)
  \mp \frac{\womega u^{z-1} }{(1-u^{z+2})} .\]
However $\E_\pm'$ can be determined from the equations of motion and a radial RG equation
for $\sigma_\pm$ can be derived.  The technical details are left to an appendix,
\ref{app:Conductivity-RG}, where 
the equation in a general dyonic background is derived 
using the techniques of \cite{H-H}, \cite{H-K-1} and \cite{BD-I}.
The result is given in equation \ref{app:Conductivity-RG} (\ref{aeq:sigma_pm'}),
\begin{align}
\pm  \womega u^{4-z}(1-u^{z+2}) u\frac{d\sigma_\pm}{d u} = &\ 4  \x (z-1)(z+2) u^{2(4-z)}(1-u^{z+2})
\left(\Slc \sigma_\pm + {\Sla}\right)^2  \nonumber   \\
                                    & \hskip 20pt - \womega^2 (\tSld^2 + \Slc^2  \x^2 u^8)\bigl(\sigma_\pm + \da\bigr)
                                                                       \bigl(\sigma_\pm + \overline \da)\label{eq:sigma_pm'} 
\end{align}
(recall that $\x=e^{-\lambda \phi_0}\frac{l^4}{r^4_h} $,  for $SL(2,{\bf R})$). Although $\chi_0$ can be absorbed into a re-definition of $b$ and $d$ for $Sl(2,{\bf R})$ this is not the case for $Sl(2,{\bf Z})$ after
Dirac quantisation).

Interpreting the classical radial equation of motion as an RG equation was suggested in
\cite{Akhmedov}, \cite{BKLT} and the relation to the $c$-theorem was studied in \cite{Freedmanetal1}, \cite{Freedmanetal2}. This version of the RG equation is obtained by expressing the radial equation of motion as a Riccati equation, \cite{Iqbal+Liu}-\cite{GTWW4}.
An alternative formulation uses the radial Hamilton-Jacobi equation \cite{BD-HJ}-\cite{BdB+H}
(for the Hamilton-Jacobi equation in the context of Lifshitz geometry see \cite{Lifshitz-Holography})
and a Wilsonian approach was developed in \cite{BKLS}-\cite{Nickel+Son} (it was argued in \cite{Sin+Yang} that the classical radial equation of motion and the Wilsonian method are equivalent).
For our purposes the Riccati equation is the most convenient form.

Ingoing boundary conditions at the event horizon give the same constraint as
in the DC case (\ref{eq:ingoing-BC}),
\[  \B_\pm  = \pm i \E_\pm,\]
at $u=1$, so $\sigma_\pm|_{u=1}=-\bar \da_h$ , but equation (\ref{eq:sigma_pm'}) already has this boundary condition encoded into it when $\womega\ne 0$,
the boundary condition at $u=1$ is no longer at our discretion for $\womega>0$,
inflowing boundary conditions are necessarily required. 

Note however that equation (\ref{eq:sigma_pm'}) implies that
$\sigma_\pm \rightarrow - \frac{\Sla}{\Slc}=-\nu$
as $\womega \rightarrow 0$, for $ 0 < u <1$, the limits $u\rightarrow 1$ and $\womega\rightarrow 0$ do not commute and
the $\womega \rightarrow 0$ limit of the AC conductivity determine by (\ref{eq:sigma_pm'})
is not the same as the DC conductivity derived in \S\ref{sec:DC-conductivity}.

\subsubsection{Analysis of the conductivity}

The solutions of the general equation (\ref{eq:sigma_pm'}) are related by 
\[ \sigma_+(-\Slb,-\Slc) =-\sigma_-(\Slb,\Slc),\]
which generalises an observation in \cite{H-H},
so the Hall conductivity  is
\[ \sigma_H=\sigma^{x y}=\frac{1}{2} (\sigma_+ + \sigma_-)=
\frac{1}{2} \bigl(\sigma_+(\Slb,\Slc) - \sigma_+(-\Slb,-\Slc)\bigr)\]
and the Ohmic conductivity  is
\[ \sigma_\Omega=\sigma^{x x}=\frac{1}{2 i} (\sigma_+ - \sigma_-)= 
\frac{1}{2 i} \bigl(\sigma_+(\Slb,\Slc) + \sigma_+(-\Slb,-\Slc)\bigr).
\]

A full analysis of the solutions of (\ref{eq:sigma_pm'}) requires numerical integration which
will not be pursued here, but certain limits are amenable to an analytic approach:

\begin{itemize}

  \item We expect a cyclotron resonance with damping, \cite{H-H,BD-I}.
An approximation to the cyclotron resonance can be found provided $\womega$ is small
and the second term on the right-hand side of (\ref{eq:sigma_pm'}) can be ignored. 
The analytic solution in this approximation is immediate,
\[ \frac {1}{\Slc(\Slc \sigma_\pm+\Sla)} = \mp \frac{4  \x (z-1)(z+2) u^{4-z}}{(4-z) \womega} + const, \qquad z\ne 4.\]
For $\sigma_+$ the boundary condition at $u=1$ is
\[\sigma_+|_{u=1}= - \overline \da_h=-\left(\frac{\tSlb-i\Sla  \x}{\tSld-i\Slc  \x}\right),\]
(recall $\tSlb = b + \Sla \lambda \chi_0$ and $\tSld = \Sld + \Slc \lambda \chi_0$).
The solution is
  \beq \sigma_+ =
    \frac{(4-z)i(\Sla  \x + i \tSlb) \womega - 4 \Sla  \x (z-1)(z+2) (1-u^{4-z})}
    {(4-z)(\tSld-i \Slc  \x) \womega + 4 \Slc  \x (z-1)(z+2) (1-u^{4-z})}, \qquad z\ne 4
\label{eq:sigma-approx}\eeq
(for $z=4$, $\lim_{z \rightarrow 4} \bigl(\frac{1-u^{4-z}}{4-z}\bigr)=-\ln u$).  

To check the range of validity of the approximation, substitute the solution (\ref{eq:sigma-approx})
into the right-hand side of (\ref{eq:sigma_pm'}) and check when the first term
dominates the second. Near $u=1$ let $u=1-\epsilon$
and $1-u^{4-z}\approx (4-z)\epsilon$ then the approximation is good provided $\x \epsilon \ll 1$
and $\x\womega \ll 1$.

  There is a resonance at
  \[ \womega_* = -\frac{4 \Slc  \x (z-1)(z+2) (1-u^{4-z})}{(4-z) (\Sld_{\chi_0}-i\Slc  \x)}.\]
  The Hawking temperature (\ref{eq:T_H_l}) was calculated in \S\ref{sec:Hawking-Temperature} in terms of $t$ in (\ref{eq:ds2r}), 
  so $e^{i \omega t}=e^{i \widetilde \omega \tilde t}$ with
  $\omega = \frac{4 \pi l \widetilde \omega T}{(z+2) \hbar}$ and
  \[ \widehat \omega = \frac{(z+2) \hbar \omega}{4 \pi T },\]
  so the resonance corresponds to
  \[ \omega_* = -\frac{16 \pi T \Slc  \x (z-1)(1-u^{4-z})}{\hbar (4-z) (\tSld -i\Slc  \x)}
    = \omega_0  - i \Gamma. \] 
  with frequency and damping
  \begin{align}
    \omega_0 & = -\frac{16 \pi \Slc \tSld  T \x (z-1)(1-u^{4-z})}{\hbar (4-z) (\tSld^2+ \Slc^2  \x^2)}
                \label{eq:omega_B}\\
    \Gamma & = \frac{16 \pi  \Slc^2  T \x^2 (z-1)(1-u^{4-z})}{\hbar (4-z) (\tSld^2+  \Slc^2  \x^2)}.
  \label{eq:Damping}\end{align}

Since $\Gamma<0$ for $z>4$ there is an instability in the system for such values of $z$ and we shall therefore assume that $1 \le z \le 4$ from now on.

Furthermore $\x \propto T^{-4/z}$ so both $\omega_0$ and $\Gamma$ vanish as $T \rightarrow 0$
but the ${\bm Q}$-factor 
  \[ {\bm Q} = \frac{|\omega_0|}{2 \Gamma} =
    \frac {1} {2  \x} \left| \frac {\tSld} {\Slc} \right|,\]
  which is independent of $u$, decreases as the temperature is decreased,
  because $\omega_0$ vanishes faster than $\Gamma$.

A similar phenomenon is seen by continuing to Euclidean time. The Hawking temperature can be derived by requiring that there is no conical singularity in the geometry in Euclidean time which in turn imposes an imaginary periodicity in real time,
$t\rightarrow t \rightarrow +\frac{i\hbar}{T}$. But a magnetic system with cyclotron frequency $\omega_B$ also has periodicity $\frac{2 \pi}{\omega_B}$ in real time and combining these gives periodicity\footnote{The periodicity in $\tilde t=\left(\frac{r_h}{l} \right)^z t$ is
  $\tilde t \rightarrow \tilde t + \frac{4\pi i l}{(z+2)}$.}
  \[t \ \rightarrow \ t + \frac{2 \pi}{\omega_B} + \frac{i\hbar}{T}.\]
  This suggests defining a complex frequency $\omega_*$ via
  \beq
    \frac{2 \pi}{\omega_*}  = \frac{2 \pi}{\omega_B} + \frac{i\hbar}{T}\quad \Rightarrow \quad
 \frac{\omega_*}{2\pi}
 = \frac{\omega_B}{2\pi}
                             \left(\frac{1-i\left(\frac{ \hbar \omega_B}{2\pi T}\right) }
                             {1+\left(\frac{\hbar \omega_B}{2\pi T}\right)^2}\right)
  \label{eq:cal-T-t}\eeq
giving frequency and damping
  \beq \omega_0  =   \frac{\omega_B }
                  {1+\left(\frac{\hbar \omega_B}{2\pi T}\right)^2},
\qquad \Gamma=  \frac{\left(\frac{\hbar \omega_B^2}{2\pi T} \right)}
{1+\left(\frac{\hbar \omega_B}{2\pi T}\right)^2} \label{eq:Euclidean-omega}\eeq
with $\bm Q$-factor
\[{\bm Q} =\frac{2\pi T}{\hbar \omega_B}.\]
The damping decreases as the temperature is lowered, as one would expect, but ${\bm Q}$ also decreases as $T$ is lowered is because the resonance frequency falls faster than the damping.

Going back to (\ref{eq:omega_B}) and (\ref{eq:Damping}) the situation near the event horizon is rather similar,
at least at small temperatures where $\x$ is large.
If the magnetic field is held fixed\footnote{In terms of  the magnetic field, using (\ref{eq:Q_m+Q_e})
  \[Q_m= B \Areah =B  \UnitA \left(\frac{r_h}{l}\right)^2
  =-\frac{c\, e^{-\lambda \phi_0}}{\eta} \quad \Rightarrow \quad c = -\eta e^{\lambda \phi_0} B  \UnitA \left(\frac{r_h}{l}\right)^2.\]
Also, since $\x=e^{-\lambda \phi_0} \left(\frac{l}{r_h}\right)^4$, $c \,\x = -\eta B \UnitA\left(\frac{l}{r_h}\right)^2   \propto \frac{\omega_B}{T^{2/z}}$.}
\[ c \,\x \propto \frac{\omega_B}{T^{2/z}}.\]
Let
\[ \frac{c\, \x}{d} = \frac{K}{2\pi}\frac{\hbar \omega_B}{T^{2/z}},\]
with $K$ a constant, and approach the event horizon by choosing\newline
\hbox{$\epsilon \approx \frac{1-u^{4-z}}{4-z}= \frac{\pi l T^{2/z}}{2 K\hbar (z-1)(z+2)}$}
  and lowering the temperature.  Then
   \beq \omega_0  =   \frac{\omega_B }
                  {1+K^2\Bigl(\frac{\hbar \omega_B}{2\pi T^{2/z}}\Bigr)^2},
\qquad \Gamma=  \frac{K\Bigl(\frac{\hbar \omega_B^2}{2\pi T^{2/z}} \Bigr)}
{1+K^2\Bigl(\frac{\hbar \omega_B}{2\pi T^{2/z}}\Bigr)^2} \label{eq:resonance_omega_0}\eeq
the same as  (\ref{eq:Euclidean-omega}) if $z=2$ and $K=1$.

\item For small $\womega \ll \womega_0$ an approximate solution of (\ref{eq:sigma_pm'}) is (see \cite{BD-I} for more details)
\[
  \sigma_+   =
  \begin{cases}
  \frac{-(  \tSlb \tSld + \Sla  \Slc  \x^2  )  + i\x }{  \tSld^2  + \Slc^2  \x^2  }, &  u=1;\\
  -\frac a c ,  &  u<1.
\end{cases}
\]
and this is an increasingly better approximation as $\womega \rightarrow 0$.

\item Exactly  at $u=1$ it is immediate from (\ref{eq:sigma_pm'}) that, for any non-zero $\womega$,
the AC conductivity is\footnote{The solution $\sigma_+ = - \da$
  is rejected because the Ohmic conductivity cannot be negative.}
\[ \sigma_+ = - \overline \da_h = \frac{-(\tSlb \tSld+ \Sla  \Slc  \x^2)  + i\x } { \tSld^2 + \Slc^2  \x^2 },\]
the same as the DC conductivity (\ref{eq:sigma-DC}).

\item As $u\rightarrow 0$, if $z<4$,
  \[ \sigma_+ \rightarrow  -\overline \da_{u=0} = -\frac{\tSlb}{\tSld}\]
for any finite non-zero $\womega$.
\end{itemize}
In summary the DC conductivity is independent of $u$, 
\beq 
  \sigma_+^{DC}  =
             \left(\frac{-\Sla  \Slc   \x^2-  \tSlb\tSld  + i\x }
               { \Slc^2  \x^2 + \tSld^2 }\right). \label{eq:sigma-IR-AC}\eeq
             In general the AC conductivity will depend on $u$ and will require numerics to
             analyse, but for $\womega\ll\womega_0$
             it reduces to a step function
\begin{align}
     \sigma_+^{AC}    = & \left(\frac{-\Sla  \Slc   \x^2-  \tSlb\tSld  + i \x }
               { \Slc^2  \x^2 + \tSld^2 }\right) , & \mbox{for}  \ u=1\\
  \sigma_+^{AC}   = &-\frac{\Sla}{\Slc } ,&  \  \mbox{for} \ 0<u<1\\
 \sigma_+^{AC}    = & -\frac{\tSlb}{\tSld}, & \mbox{for}  \ u=0.\label{eq:sigma-IR-DC}
\end{align}

\section{Temperature flow in the infra-red\label{sec:T-flow}}

Since the infra-red conductivity is $\sigma_+ = - \overline \da|_{u=1}$ the conductivity transforms the same way as the dilaxion under an $Sl(2,{\bm R})$ transformation\footnote{While this is the case at the event horizon it is not true for $u<1$ (see appendix \S\ref{app:modular}), but in this subsection we focus on $u=1$.}
\beq \sigma_\pm = -\overline \da_h \quad \rightarrow \quad
  - \left(\frac{ a \overline \da_h + b}{c \overline \da_h + d}\right)= \left(\frac{ a \sigma_\pm - b}{-c \sigma_\pm + d}\right).\label{eq:gamma-sigma}\eeq
Invoking the Dirac quantisation condition in the bulk
  \beq\sigma= -\overline \da|_{u=1}=
\frac{(\ib + \ia \lambda \chi_0) \Theta^{4/z} - i \ia} {i \ic-(\id + \ic \lambda \chi_0) \Theta^{4/z} }\label{eq:sigma_horizon}\eeq 
at the event horizon (we can drop the $\pm$ subscript when $\womega=0$, since then $\sigma = \sigma_+$ and $\overline \sigma=-\sigma_-$ and the only difference between $\sigma_+$ and $\sigma_-$ is the sign of the Hall conductivity $\sigma_H$,
in the following $\sigma$ is $\sigma_+$).
In the purely electric case, $\ia=\id=1$, $\ib=\ic=0$,
\beq \sigma  =-\lambda  \chi_0  +  i \Theta^{-4/z} \label{eq:pure-electric},
\eeq
with $- \frac 1 2  \le \lambda  \chi_0  \le \frac 1 2 $, and
and the Ohmic conductivity diverges as $T\rightarrow 0$, which would be normal behaviour of a conductor
in the absence of impurities. 
When $T$ is increased the Ohmic resistance grows as $T^{4/z}$, giving a power law with exponent $1\sim 4$ for
$4\ge z \ge 1$. Conversely in the purely magnetic case, $\ia=\id=0$, $\ib=-\ic=1$,
\beq \sigma  = \frac {\lambda  \chi_0  \Theta^{8/z} + i \Theta^{4/z}}
{1 + \lambda^2  \chi_0 ^2 \Theta^{8/z}} \label{eq:pure-magnetic},
\eeq
both the Ohmic and the Hall conductivities tend to zero  as $T\rightarrow 0$. 

$Sl(2,{\bm Z})$ is generated by ${\bm S}:\sigma \rightarrow -1/\sigma$ and ${\bm T}:\sigma \rightarrow \sigma +1$. In particular $\Theta=1$, $ \chi_0 =0$ is a fixed point under $S$-duality,
with critical temperature $T_*$, and in the dual phase $\sigma \rightarrow 0$ as $T\rightarrow 0$ for a background with a magnetic charge but no electric charge.   This fixed point is similar
to the one discussed in \cite {Peskin} \cite{Dasgupta+Halperin}, in \cite{Fisher-Lee} \cite{Fisher} it was interpreted as being
due to a superconductor-insulator phase transition arising from
Bose condensation of vortices, in which the insulating phase is a Hall insulator.

For $- \frac 1 2  \le  \lambda  \chi_0 \le \frac 1 2 $, $S$-duality places the critical temperature on the boundary of the fundamental domain, where $|\sigma|=1$,
at $\Theta = \left(\frac{1}{1-\lambda^2  \chi_0 ^2}\right)^{z/4}$, so
the maximum critical temperature is at $\lambda \chi_0 =\pm \frac 1 2$ where $T= \bigl(\frac 4 3\bigr)^{z/4} T_*$. 

However we now have an apparent paradox. In the classical solution the metric is not
affected by $Sl(2,{\bm R})$ transformations, in particular $r_h$ and hence $T$
are invariant. They should therefore be invariant under $\Gamma(1)$ and yet ${\bm S}$
interchanges large and small temperatures.  To resolve this we note that $T\rightarrow \infty$ is not accessible to the classical solution as we must constrain $T\ll T_{Pl}$ and we
consider two possible strategies:
\begin{itemize}
\item Eliminate ${\bm S}$ and only allow sub-groups of $\Gamma(1)$ that do not contain ${\bm S}$.

\item  Retain ${\bm S}$ but keep $T\le T_*$.

\end{itemize}

We shall examine these two possibilities in turn, exploring the second one first.

\subsection{Retain ${\bm S}$ but keep $T\le T_*$\label{sec:Gamma1}}

\begin{figure}[h!]
    \centerline{\includegraphics[width=12cm]{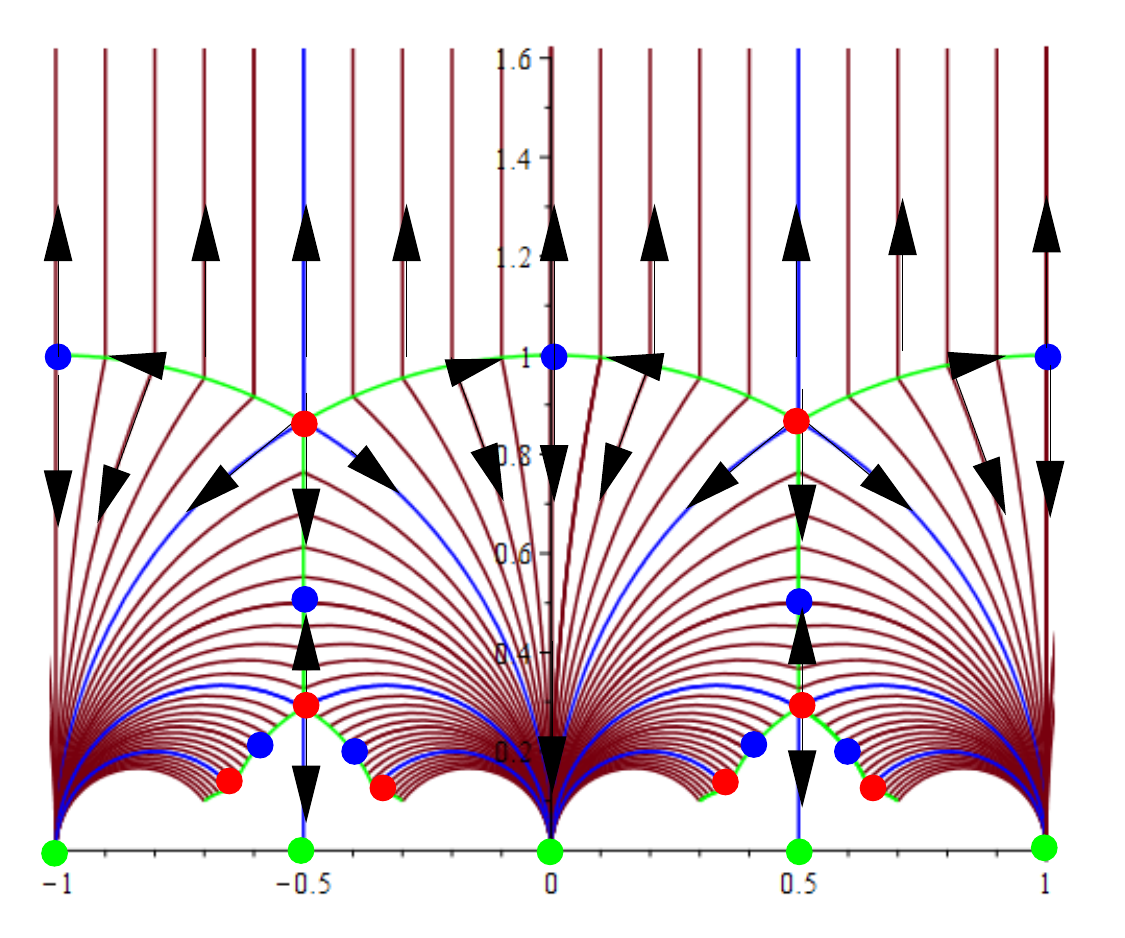}}

    \begin{caption} { \small Temperature flow of the conductivity on the event horizon
        obtained from tiling the upper-half conductivity plane, replicating
        the fundamental domain using $\Gamma (1)$ symmetry (arrows in direction of decreasing temperature).
    The green lines indicate phase boundaries, 
    blue circles are saddle points and the red circles are purely repulsive fixed points.
    Green circles are purely attractive fixed points as $T\rightarrow 0$ (there are attractive fixed points at every rational number on the real axis, but only half-integral values are
    shown in the figure).
    The vertical strip $-\frac 1 2 \le \sigma^{x y} \le \frac 1 2$ above the green circular arc between $e^{-i\pi/3}$ and $e^{i\pi/3}$, bounded by the vertical blue lines, is a fundamental domain
    of $\Gamma(1)$.}
  \label{fig:Sl2Z-naive-flow}
\end{caption} 
\end{figure}

With the superconductor-insulator transition in mind use the electric solution ($\ia = \id =1$, $\ib = \ic =0$, $\sigma=-\lambda \chi_0+i\Theta^{-4/z}$, $-\frac 1 2 \le \chi_0 \le \frac 1 2$)
above the green semi-circular arc bounding the lower edge of the fundamental domain in figure \ref{fig:Sl2Z-naive-flow}, which flows up to the superconductor as
$T\rightarrow 0$, and use the magnetic solution ($\ia = \id =0$, $-\ib = \ic =1$,
$\sigma=\frac{\lambda \chi_0 \Theta^{8/z} + i\Theta^{4/z}} {1+ \lambda^2 \chi_0^2 \Theta^{8/z}}$)
below the green arc, flowing to the Hall insulator as $T\rightarrow 0$. 
  This point of view resonates with the introduction of the complex frequency in (\ref{eq:cal-T-t}). Continuing time into the complex plane, complex periodicity suggests defining
 \beq
 {\cal T}= T_* \left(\frac{2 \pi}{\omega_B} + \frac{i\hbar}{T},\right)
\eeq
as a Teichm\"uller parameter for a torus.
There is then an $Sl(2,{\bm Z})$ action
\[ {\cal T} \rightarrow \frac{ \ia {\cal T} +\ib}{\ic {\cal T} + \id}\]
on this temporal torus. If $T \ll \hbar \omega_B$ (a necessary condition to access the hierarchy of
phases in the QHE) then ${\cal T} \approx \frac{i \hbar}{T}$ and ${\bm S}:{\cal T} \rightarrow -\/{\cal T}$
has a fixed point at the critical temperature $\Theta=1$.

 The idea is then that different filling fractions $\frac {\ia} {\ic}$ label different quantum phases and the flow lines in the fundamental domain $(\ia=\id=1$, $\ic = \ib =0)$ are mapped between different phases by the action of $\Gamma(1)$.
 Temperature flow lines, arising from varying $r_h$ in the fundamental domain,
 with \hbox{$-\frac 1 2 \le \lambda \chi_0 \le \frac 1 2$} fixed, are shown in figure \ref{fig:Sl2Z-naive-flow}. As the temperature is reduced there are repulsive fixed points at $\sigma=e^{i \pi/3}$, and its images under $\Gamma(1)$ (red circles);
 attractive fixed points at all rational numbers on the real axis (half-integral values are shown in green), as well as at $Im(\sigma)\rightarrow i \infty$; and saddle points  at $\sigma=i$, and its images (blue).  The flow looks a little confused on the green lines, which are phases boundaries,  but this will be resolved momentarily.

 Of course the other flow direction could have been chosen, nothing in the mathematical analysis dictates which direction to use: the direction is chosen on physical grounds, to make the superconductor in the electric phase attractive as $T\rightarrow 0$.  Different classical bulk solutions are being used to represent different phases of the 2-dimensional quantum system.

\subsubsection{Gradient flow for $\Gamma(1)$}

We now take up the suggestion in \cite{L+R:anti-holomorphic1} that this flow might be derivable as gradient flow with an anti-holomorphic potential.  
In a conductor or a superconductor the temperature flow of the conductivity
is driven by the fact that there is an underlying electron coherence length that
depends on the temperature and the conductivity depends on the temperature via the
coherence length.
In general we can define a flow function resulting from changing the electron coherence length via a change in the temperature as
\[ T\frac{d \sigma} {d T}= T \frac{d \xi}{d T} \frac{ d \sigma}{ d \xi}. \]
Changes in the coherence length length generate a vector field in the conductivity plane
\[ \beta(\sigma, \bar \sigma)=\xi \frac{d \sigma}{d \xi} \]
and, under a modular transformation\footnote{The sign of $\ib$ and $\ic$ have been changed compared to (\ref{eq:gamma-sigma}) to avoid cluttering subsequent formulae with minus signs --- this is still a modular transformation, it is equivalent to changing the sign of $Re(\sigma)$.} $\sigma \rightarrow \frac{\ia \sigma +\ib}{\ic \sigma + \id}$,
\beq \beta(\sigma, \bar \sigma) \rightarrow \frac{\beta(\sigma, \bar \sigma)}{(\ic\sigma + \id)^2}.
\label{eq:beta-transformation}\eeq
If this flow is derivable from a potential $\overline V(\sigma,\bar \sigma)$ then
\beq \beta(\sigma, \bar \sigma) = G^{\sigma \bar \sigma} \partial_{\bar \sigma}
\overline V(\sigma,\bar \sigma)\label{eq:beta-def}\eeq
with $G_{\sigma \bar \sigma}$ a hermitian metric on the conductivity plane. 
There is of course a natural candidate for an $Sl(2,{\bm R})$ invariant metric
and that is
\[ G_{\sigma \bar \sigma} = \frac{1}{(Im  \sigma)^2}.\]
The proposal in \cite{L+R:anti-holomorphic1} is that the potential be
anti-holomorphic $\overline V(\bar \sigma)$.
Under a modular transformation
\[ Im \sigma \rightarrow \frac{Im \sigma}{|\ic \sigma +\id|^2} \]
so
\[ G^{\sigma \bar \sigma} \rightarrow \frac{G^{\sigma \bar \sigma}}{|\ic \sigma +\id|^4} \]
and (\ref{eq:beta-def}) will have the transformation property (\ref{eq:beta-transformation})
if
\beq \overline V(\bar \sigma) \rightarrow (\ic \bar \sigma +\id)^2 \,\overline V(\bar \sigma),\eeq
{\it i.e.}  $1/V(\sigma)$, the complex conjugate of $1/\overline V(\bar \sigma)$, is a modular form
of weight $-2$.

This severely restricts the form of the potential as there is a theorem \cite{Rankin} that any modular form
$\Phi_{-2}(\sigma)$ of weight -2 can be written in terms of Klein's $J$-invariant as\footnote{Relevant properties of modular forms are summarised in appendix \S\ref{app:modular-forms} for convenience.}
\[ \Phi_{-2}(\sigma)=  \frac{P(J)}{Q(J)}\frac{1}{J'}\]
where $P(J)$ and $Q(J)$ are polynomials in $J$ and $J'=\frac{d J}{d\sigma}$.
We shall therefore investigate using $V(\sigma)=\{\Phi_{-2}(\phi)\}^{-1}$
as a potential.

Under this assumption the flow commutes with the $\Gamma(1)$ action and fixed points of
$\Gamma(1)$, {\it i.e.} points in the $\sigma$-plane that are left invariant by
at least one non-trivial element of $\Gamma(1)$, are necessarily fixed points of the flow,
\[ \sigma_* = \frac{\ia \sigma_*+\ib}{\ic \sigma_* +\id} \quad \Rightarrow \quad
  \beta(\sigma_*,\bar \sigma_*)=\frac{\beta(\sigma_*,\bar \sigma_*)}{(\ic \sigma_*+\id)^2}
  \quad \Rightarrow \quad
  \beta(\sigma_*,\bar \sigma_*)=0 \quad \mbox{or} \quad \infty.\]
There are three points in the fundamental domain that are left invariant by some element of $\Gamma(1)$:
$\sigma_{1,*}=i\infty$, $\sigma_{2,*}=e^{i\pi/2}$ and $\sigma_{3,*}=e^{i\pi/3}$,  with $J$ taking values $+\infty$, $1$ and $0$, respectively at these three points. In fact $J$ is real and monotonically increasing from $0$ to $1$ as
$\sigma$ goes from $e^{i\pi/3}$ to $e^{i\pi/2}$, along the unit circular arc, and then increases
from $1$ to $\infty$ along the vertical line as $\sigma$ runs from $i$ to $i \infty$.

There can be more fixed points though, a zero of $Q(J)$ would give $\beta = 0$
and a zero of $P(J)$ would give a divergent $\beta$. The simplest assumption is that there
are no fixed points of the flow other than the fixed points of $\Gamma(1)$. 
A property of $J$ is that it takes all possible complex values once and only once in the fundamental
domain of $\Gamma(1)$, so if $P(J)$ and $Q(J)$ are to have no zeros other than at
$e^{i\pi/3}$, $i$ and $i\infty$ it must be the case that
\[ \frac{{Q(J)}}{{P(J)}}= -\overline C J^m (J-1)^n\]
  for some pair of if integers $m$ and $n$, with $C$ a constant.
  Furthermore $m$ and $n$ are restricted by the form of the classical solution of the equations of
  motion.

  Consider the purely electric solution (\ref{eq:pure-electric}) with $\chi_0=0$ and
\[ \sigma = i \Theta^{-4/z}, \]
above the fixed point $\Theta =i$ with $\Theta\le 1$.
The metric on this line is
\[G_{\sigma \bar \sigma}=\Theta^{8/z}\]
so
\[ \beta =  -\frac{C \overline J^m (\overline J-1)^n}{\Theta^{8/z}}\frac{ d \,\overline J}{d \bar \sigma}
  = \xi\frac{d \sigma}{d \xi}.\]

Now consider the three types of fixed point:
\begin{itemize}
  \item $\bm{\sigma_{*,1}, \sigma \rightarrow i\infty}$: in terms of $q=e^{i \pi \sigma}$, $J$ has the small $q$ expansion
\[ J = \frac{1}{1728}\Bigl(q^{-2} + 744 + 196884 q^2 + 21493760\,{q}^{4}  +O \bigr( {q}^{6} \bigr) \Bigr)\]
so as $\Theta \rightarrow 0$
\[ \beta \ \approx \ -\frac{2 \pi i C \Theta^{-8/z}}{(1728)^{m+n+1}} e^{\frac{2\pi(m+n+1)}{ \Theta^{4/z}}}\]
giving
\[ \xi \frac{ d \Theta^{-4/z}}{d \xi} \ \approx \
  -\frac{2 \pi  C \Theta^{-8/z}}{(1728)^{m+n+1}}  e^{\frac{2\pi (m+n+1) }{\Theta^{4/z}}}\]
(clearly $C$ must be real for $\xi$ to be real).

If the $\beta$-function is to be analytic as $T\rightarrow 0$, $m$ and $n$ must
be restricted by the constraint $m+n+1 = 0$.
This gives
\beq \xi \approx \xi_0 \exp\left({\frac{\Theta^{4/z}}{2\pi C}}\right)\ \rightarrow \ \xi_0 \quad \mbox{as}\quad  \Theta \rightarrow 0 \label{eq:BCS-correlation-length}\eeq
and
the coherence length is a non-zero constant as
$T\rightarrow 0$.  

\item $\bm{\sigma_{*,2}}=i$: for $\sigma = i + \epsilon$ with $\epsilon$ small
\[ J \approx 1 - k \epsilon^ 2,\]
with $k = \frac{3}{64 \pi^4} \left\{\Gamma \left(\frac {1} {4}\right)\right\}^8\approx14.37$,
so
\[ \beta \approx  2 C k  \bar \epsilon (-k\bar \epsilon^2)^n,\]
since $G^{\sigma \bar \sigma}=1+O(\epsilon)$.
Demanding that $\beta$ is finite at $\sigma=i$ restricts us to $n\ge 0$.
Now let  ${\tt t}=1-\Theta$ be the reduced temperature near the critical point, with $0\le {\tt t} \le 1$, then
\[ \bar \epsilon = -\frac{4 i {\tt t}}{z}\]
and
\[\beta  \approx   -2  i C k^{n+1}\left(\frac{4  {\tt t}}{z}\right)^{2n+1}.   \] 
Together with 
\[  \beta   = \xi \frac{d {\tt t}}{d \xi} \frac{d \sigma} {d {\tt t}} =  \frac{4 i} {z}\left(\xi \frac{d {\tt t}}{d \xi}\right)
\]
this gives 
\begin{align*}
         \frac{1}{\xi}\frac{d \xi}{ d {\tt t}} & \approx -\frac{1}{2 C k^{n+1}} \left(\frac{z}{4}\right)^{2 n}  {\tt t}^{-(2 n+1)}\\
  \Rightarrow \quad  \ln\xi & \approx
  \begin{cases}
    \frac{1}{4 n  C } \left(\frac{z}{4}\right)^{2 n} k^{-(n+1)}{\tt t}^{- 2 n}, & n\ne 0;\\ 
  -\frac{1}{2 k C }  \ln {\tt t}, & n=0.
  \end{cases}
\end{align*}
$n=0$ gives the scaling behaviour 
\[ \xi \approx A {\tt t}^{-\nu'}\]
with $A$ a constant and critical exponent
\[ \nu' = \frac{1}{2 k C}= \frac{32 \pi^4}{3 C \bigl\{\Gamma(\frac{1}{4})\bigr\}^8 }.\] 

\item $\bm{\sigma_{*,3}=e^{i\pi/3}}$: what of the third fixed point at $\sigma_{3,*}=e^{i\pi/3}$?
Near $e^{i\pi/3}$, with \hbox{$\sigma = e^{i \pi/3} +\epsilon'$}, $J$ vanishes as
\[J\approx   - i k' (\epsilon')^3\]
with $k'$ a positive constant.\footnote{$k' = \frac{2^8}{(\sqrt{3}\pi)^3}
  \left\{ K\left( \sin\left( \frac{\pi}{12} \right)\right)\right\}^6\approx 26.47$,
  with $K$ the elliptic integral of the second kind, but we shall not need its explicit value.}
The metric takes the value $G_{\sigma \bar \sigma} = \frac{4}{3}$ giving
\[ \beta \approx \frac{9  C}{4} (-1)^{n+1}\bigl(e^{i \pi/2} k'\bigr)^{m+1} (\bar \epsilon\,')^{3 m+2}.\]

Demanding that $\beta$ is finite at $\epsilon'=0$ would constrain $m$ to be positive, but requiring
$\beta$ to be analytic at all three fixed points gives three incompatible conditions,  $m+n=-1$, $n\ge 0$ and $m\ge 0$.
We do not yet have a physical interpretation of $\epsilon'$ so we shall keep $m+n+1 = 0$ and $n\ge 0$ and allow $m<0$.  The marginal case is $n=0$, $m=-1$, in which case there is a simple pole in $\beta$ at $\sigma_{*,3}$.

On the unit circular arc from $\sigma_{*,2}$ to  $\sigma_{*,3}$,
$\sigma = e^{i(\frac{\pi}{3} + \psi)}$
with $0 \le \psi \le \frac{\pi}{6}$, $J$ is real and lies in the range  $1 \ge J \ge 0$.
Close to $e^{i\pi/3}$, where $\psi \ll 1$,
\[ \epsilon' \approx e^{5 i \pi/6} \psi\]
and
\[ \xi \frac{ d \psi}{d \xi}  \approx -\frac{9 C }{4 \psi}
  \quad \Rightarrow \quad \xi \approx \tilde \xi_0 e^{-\frac{2\psi^2}{9 C}}.\]

\end{itemize}
In summary the behaviour of $\xi$ near the fixed points is
\beq \xi \sim
  \begin{cases}
    \xi_0 \exp\left(\frac{\Theta^{4/z}}{2 \pi C}\right), & \Theta \rightarrow 0;\\
    A {\tt t}^{-\nu\,'}, & \Theta \rightarrow 1, \ \mbox{with} \ \nu\,' =\frac{32 \pi^4}{3 C \bigl\{\Gamma(\frac{1}{4})\bigr\}^8 };\\
    \tilde \xi_0 \exp \left(-\frac{2 \psi^2}{9 C}\right), & \sigma =e^{i (\pi/3+\psi)} \rightarrow e^{i\pi/3}. \label{eq:Gamma1-beta-function}  \end{cases}\eeq
The classical solution in the bulk does not give us $\psi$ as a function of $T$,
but a consistent picture emerges if $\psi$ is monotonic in $T$ and $T$ increases monotonically from $T_*$ to infinity
along the circular arc connecting the two fixed points $e^{i \pi/2}$ and $e^{i\pi/3}$.
The hypothesis of gradient flow has resolved the ambiguity across the green arc of points in figure \ref{fig:Sl2Z-naive-flow} and replaced it with a separatrix between the two adjoining phases.

\begin{figure}[h!]
\centerline{\includegraphics[width=10cm]{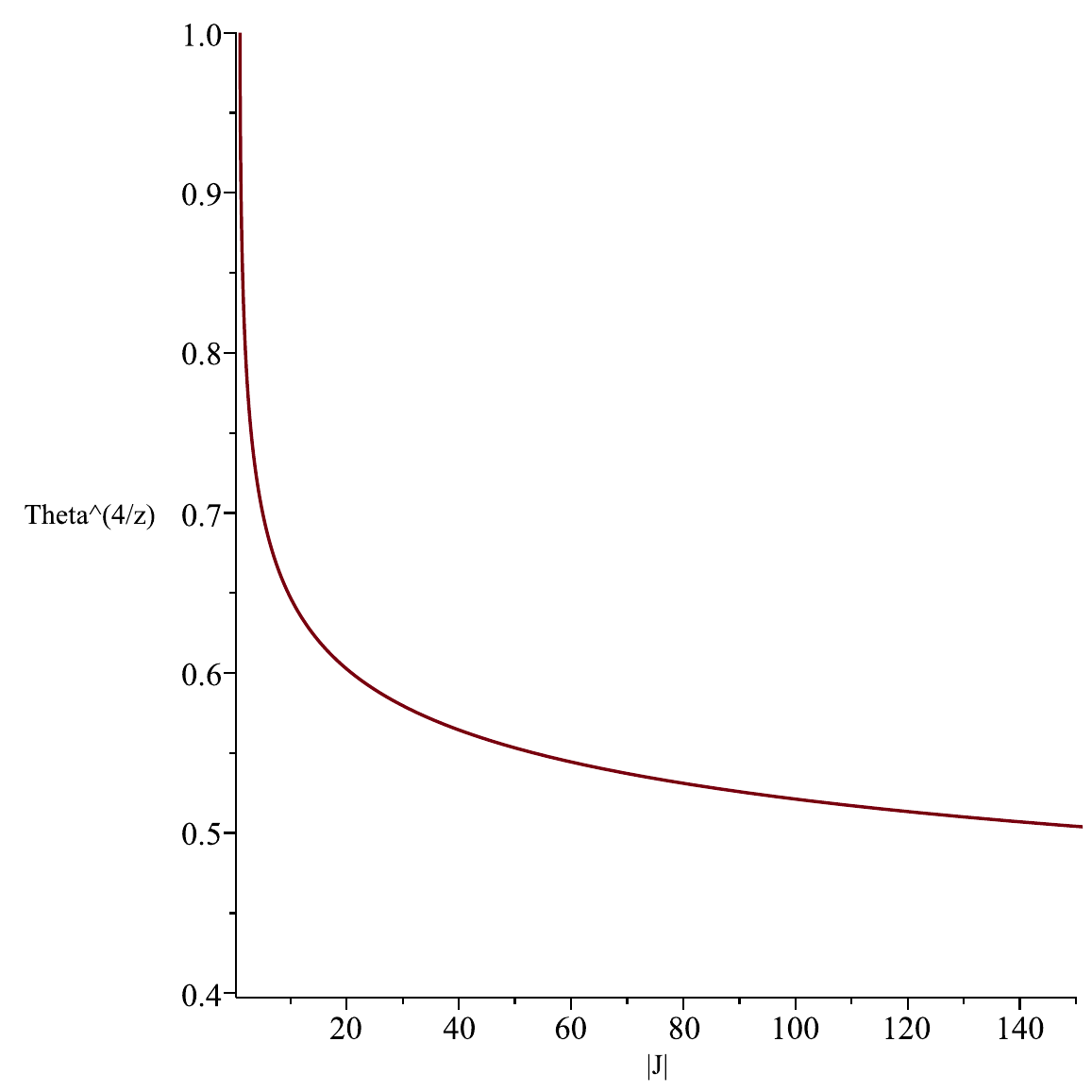}}
\begin{caption}
  {\small Temperature as a function of $|J|$ for $|J|\ge 1$ along the imaginary axis, with $\Gamma(1)$ flow.
    $\Theta$ drops monotonically from unity
   at $|J|=1$ to zero as $|J|\rightarrow \infty$.
  Asymptotically $\Theta^{4/z}\approx \frac{2\pi}{\ln|J|}$. This assumes that the classical solution gives the correct temperature dependence along the imaginary axis. }
\label{fig:sigma_v_J}\end{caption} 
\end{figure}

\begin{figure}[h!]
\centerline{\includegraphics[width=14cm]{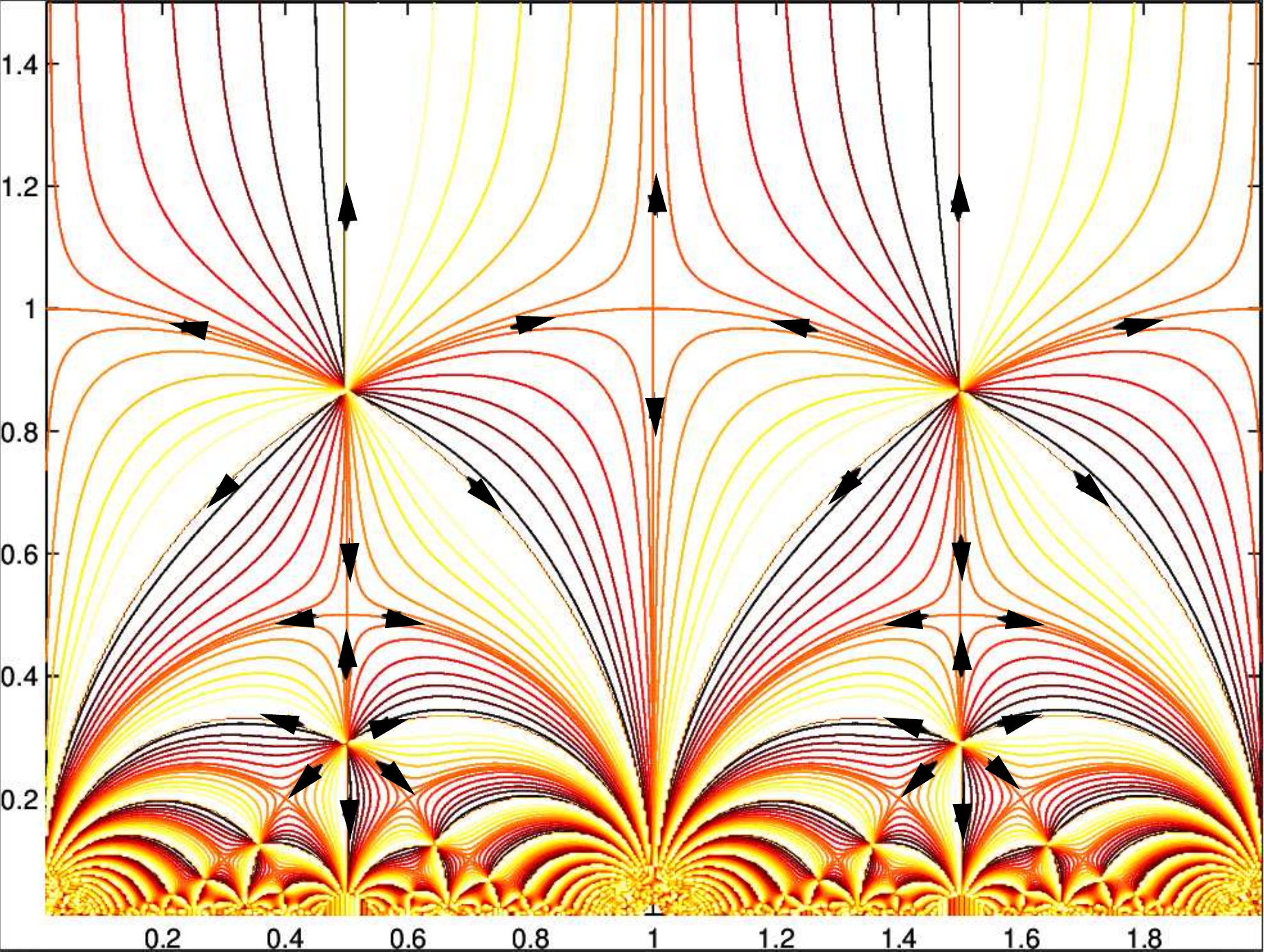}}
\begin{caption} { \small Temperature flow of the conductivity on the event horizon
    for $\Gamma (1)$ symmetry, assuming it is generated by varying $|J|$, keeping $\arg(J)$ fixed.
    The diagram is obtained simply by plotting lines of constant $\arg(J)$, which are represented by different colours from red ($\Psi=0$) through black ($\Psi=\pi$) to yellow.}
\label{fig:Sl2Z-full-flow}\end{caption} 
\end{figure}

 On the upper segment of the
imaginary axis, $\sigma=i \Theta^{-4/z}$ for $\Theta$ in the range $1 \ge \Theta \ge 0$ while
$J$ is real and lies in the range $1\le J(\sigma)<\infty $ and we can invert the explicit formula for $J(\sigma)$, (\ref{eq:J-def}) in appendix \ref{app:modular-forms})
to plot $\Theta$ in this range as a function of $J$ (this is shown in figure
\ref{fig:sigma_v_J}).
Now the temperature, and hence $\Theta$, depends only on the metric and not
on any $Sl(2,{\bm R})$ transformation of the classical solution, and so $\Theta$ should be modular invariant. Suppose therefore that $\Theta$ can be extended to a real function $\Theta(J,\overline J)$
over the whole of the fundamental domain, and by extension using $\Gamma(1)$ to the whole upper-half complex plane.  If $\Theta$ is a monotonic function of $|J|$ everywhere then flow lines can be obtained by varying $|J|$, keeping its argument $\Psi$ fixed.
With $J=|J|e^{i \Psi}$ and
$m=-1$, $n=0$
\[ \xi \frac{d \sigma}{d \xi}
= - C  (\sigma^{x x})^2 \frac{d \ln \overline J}{d \overline \sigma} =
  - \frac{C}{2} \frac{d (\ln |J|^2)} {d \overline \sigma}
  + i C  (\sigma^{x x})^2 \xi \frac{d \Psi}{d \xi}.\]
Now
\[ \Psi = \tan^{-1} \left\{ \frac{i(\overline J - J)}{(\overline J + J)} \right\}
 \qquad \Rightarrow \qquad \frac{d \Psi}{d \sigma}=-\frac{i}{2 J}\frac {d J}{d \sigma}\]
and
\begin{align*} \xi \frac{ d \Psi}{ d \xi}
  & = \xi \frac{d \sigma}{d\xi} \frac{d \Psi}{d\sigma}
    + \xi \frac{d \,\overline \sigma}{d\xi} \frac{d \Psi}{d\,\overline\sigma}\\
  &=
  -C (\sigma^{x x})^2\left\{
    \frac{d \ln \overline J}{d \overline \sigma}
    \left( - \frac{i}{2  J} \right)\left( \frac{d J}{d\sigma}\right)
        +
        \frac{d \ln  J}{d \sigma}
        \left( \frac{i}{2\overline J} \right)\left(\frac{d \overline J}{d \overline \sigma}\right)
      \right\}\\& = 0\end{align*} 
    if $C$ is real.
    Thus for real $C$ the argument of $J$ is indeed constant along the flow lines.
    Furthermore
    \begin{align*}
      \xi \frac{d |J|^2}{d \xi} & =
      \xi \frac{d \overline \sigma} {d \xi} \frac{d |J|^2}{d \overline \sigma}
                                 +\xi \frac{d\sigma} {d \xi} \frac{d |J|^2}{d  \sigma}\\
           &= - C (\sigma^{x x})^2
  \left\{\frac{d \ln J}{d \sigma}\frac{d |J|^2}{d \overline \sigma}
      + \frac{ d \ln \overline J}{d \overline \sigma} \frac{d |J|^2}{d \sigma}
                                 \right\}\\
     & = -2 C \left( \sigma^{x x} \left| \frac{d J}{d\sigma}\right|\right)^2\\
        \Rightarrow \qquad
       \xi \frac{d \ln |J|}{d \xi}
           & = - C \left( \sigma^{x x} \left|\frac{ d \ln |J|}{d \sigma} \right|\right)^2
  \end{align*}
  hence $|J|$ is indeed a monotonically decreasing function of $\ln \xi$.
    The temperature flow can therefore be visualised as lines of constant $\Psi$ in
  the conductivity plane and these
  are plotted in figure \ref{fig:Sl2Z-full-flow} (this figure is obtained using the logic in \cite{Crossover,anti-holomorphic-v-holomorphic}).\footnote{$\xi$ need not be precisely the same as electron coherence length for deriving the temperature flow diagram, all that is necessary is that $T(\xi)$ is a monotonic function of $\xi$ in the fundamental domain and same diagram would ensue.}

$J$ is real on the segment of the
unit circle centered on the origin that connects these two fixed points
and decreases from 1 at $\sigma_{*,2}$ to to 0 at $\sigma_{*,3}$, as $\Theta$ rises
  from $1$ to infinity. Thus the full temperature range $0\le \Theta < \infty$ is re-instated
  in a self-consistent flow.  The fixed point at $e^{i\pi/3}$ is a sink in the high
  temperature direction  
  that takes the theory out of its domain of validity as the temperature is increased,
  in the AdS/CMT paradigm this would suggest that quantum effects become significant in the bulk and the 2-D quantum system becomes weakly coupled at $\sigma^{i\pi/3}$. 

   It is not necessary to know the
   function $\Theta(J,\overline J)$ explicitly in order to plot figure \ref{fig:Sl2Z-full-flow}, it can be any monotonic function of $|J|$.  The classical solution only gives $\Theta$ for $\arg(J)=0$,
and even then only for $J\ge 1$. It is a strong assumption that $\Theta(|J|,\Psi)$ is monotonic
in $|J|$ everywhere and that the temperature flow is obtained by varying $|J|$, but the picture that emerges under these assumptions is at least consistent.  In the next section similar assumption are made for
the sub-group $\Gamma_0(2)\subset \Gamma(1)$ and give very good agreement with experimentally measured flows.

  \subsection{Eliminate ${\bm S}$\label{sec:Gamma02}}

  Another way of obtaining a consistent temperature flow is to eliminate ${\bm S}$ and consider
  a subgroup of $\Gamma(1)$.  For example we could consider the set of elements generated by repeated  application of $\bm{S T S}$ and $\bm T$, but \hbox{$(\bm{S T})^3=1$} so $(\bm{S T S}){\bm T} (\bm{S T S})= \bm{S}$ and
  $\bm S$ re-appears. However $\bm{S T}^2 \bm{S}$ and $\bm T$ do generate a group --- the  subgroup $\Gamma_0(2)$ of $\Gamma(1)$,
  and this is one option, but there are others \cite{1917}, \cite{Witten-Sl2Z} (appendix \S\ref{app:modular-forms} collects together
  some relevant facts about level 2 subgroups of $\Gamma(1)$ and their modular forms).
  
\begin{figure}[h!]
\centerline{\hspace{40pt}\includegraphics[width=12cm]{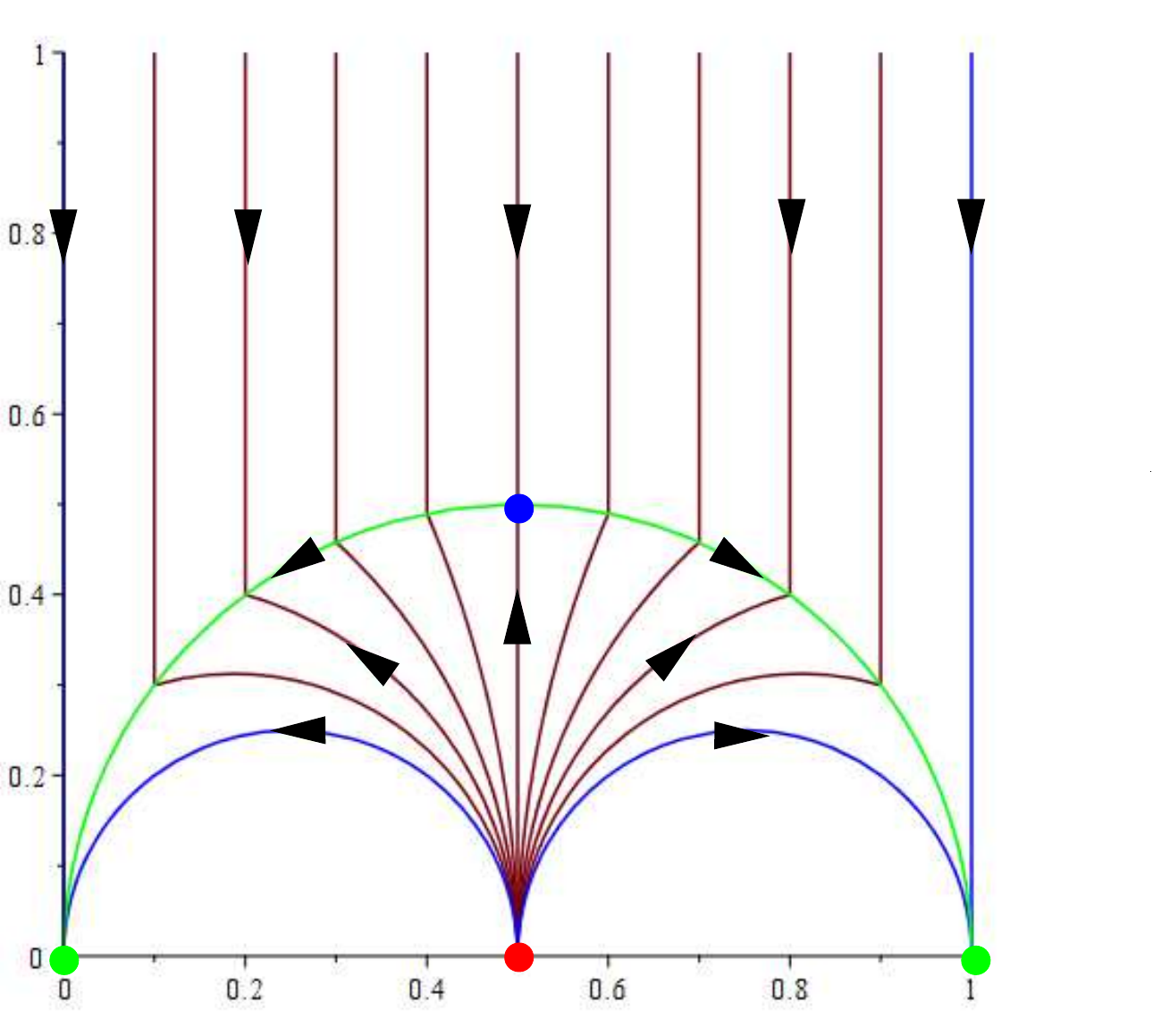}}


\begin{caption} { \small Temperature flow of the conductivity on the event horizon
    for the group $\Gamma_0(2)$. The fundamental domain can be taken to be the vertical strip
    $0\le \sigma^{x y} \le 1$ above the green semi-circular arc of radius one-half, spanning $0$ and $1$, the pattern repeats for $\sigma^{x y} \rightarrow \sigma^{x y} + n$ for any integer $n$.
    The temperature flow of the purely magnetic solution in the fundamental domain is shown
    together with one other copy obtained by applying the $\Gamma_0(2)$ transformation
  $\bpm 1 & -1 \\ 2 &-1 \epm \in \Gamma_0(2)$ which leaves $\sigma_*=\frac{1}{2} (1 + i)$ fixed. }
\label{fig:Gamma02-naive-flow}\end{caption} 
\end{figure}

The group $\Gamma_0(2)$ has two fixed points on the imaginary axis, at $\sigma=i\infty$ and $\sigma=0$,
but the $\Gamma(1)$ fixed point  at $\sigma_{2,*}$ is gone.
The temperature flow diagram generated by taking the purely magnetic solution 
\[\sigma=-\lambda \chi_0 + i \Theta^{4/z} \]
in the fundamental domain of $\Gamma_0(2)$,
with $-1\le\lambda  \chi_0 \le 0$ fixed, and mapping it around with elements of $\Gamma_0(2)$ is shown in figure \ref{fig:Gamma02-naive-flow}.
$\Gamma_0(2)$ has a third fixed point at $\sigma_*=\frac 1 2 (1 + i)$,
left invariant by $\bpm 1 & -1 \\ 2 & -1 \epm $ which reflects the green arc bounding the fundamental domain in figure \ref{fig:Gamma02-naive-flow} about the vertical line $\sigma^{x y}=\frac 1 2$.  Again there is confusion along the green semi-circular arc that constitutes the
lower boundary of the fundamental domain in the figure, and modifying the flow to a gradient flow resolves this to a separatrix with a saddle point at $\sigma=\frac{1+i}{2}$.

\subsubsection{Gradient flow for $\Gamma_0(2)$}
For $\Gamma_0(2)$ there is a parallel theorem about modular forms to
that used for $\Gamma(1)$ in \S\ref{sec:Gamma1}, \cite{Rankin}: any modular form of weight -2 for $\Gamma_0(2)$ can be written as
\[ \widetilde \Phi_{-2}(\sigma) = \frac{\widetilde P(f)}{\widetilde Q(f)}\frac{1}{f'}\]
where $f(\sigma)$ is the $\Gamma_0(2)$ invariant function
\[ f(\sigma) = - \frac{1}{256 q^2} \prod_{n=1}^{\infty}\frac{(1-q^{4n-2})^8}{(1+q^{2 n})^{16}},\]
with $q = e^{i \pi \sigma}$ and $\widetilde P(f)$ and $\widetilde Q(f)$ polynomials in $f$.  So we can investigate $\overline {\widetilde V}(\bar \sigma)=\{\overline {\widetilde \Phi}_{-2}\}^{-1}$ as a potential. On the vertical line in figure \ref{fig:Gamma02-naive-flow}, associated with the magnetic solution with $\chi_0$, 
$\sigma= i\Theta^{4/z}$, $q=e^{-\pi\Theta^{4/z} }$ is real and hence $f$ is real
on this line, it is in fact negative and runs monotonically down from $0$ to $-\infty$ as $\Theta$
increases from $0$ to $\infty$.
Since $f$ is invariant under $\Gamma_0(2)$ transformations it is real and negative
on the blue semi-circles in figure \ref{fig:Gamma02-naive-flow}, which images of the imaginary axis under $\Gamma_0(2)$.

However $\Gamma_0(2)$ has another fixed point at $\sigma_*=\frac{1+i}{2}$, and its images, and $f=\frac{1}{4}$ at $\sigma_*$.  Just as $J$ does for $\Gamma(1)$, $f$ takes all complex values once and only once in the
fundamental domain of $\Gamma_0(2)$.
With the same assumption as in section \S\ref{sec:Gamma1}, that there are no fixed points
of $\beta$ other than the fixed points of $\Gamma_0(2)$, the form of $\widetilde P(f)$ and $\widetilde Q(f)$ are then constrained so that
\[ \frac{\widetilde Q(f)}{\widetilde P(f)} = \overline C f^m \left( \frac 1 4 - f\right)^n\]
and
\[ \beta = C \Theta^{8/z}\bar f^m \left(\bar f-\frac{1}{4}\right)^n \frac{d \bar f}{d \bar \sigma}\]
for the purely magnetic solution, with $C$ a constant.
The reasoning now follows along the same lines as for $\Gamma(1)$, except the purely
magnetic solution is used on the imaginary axis bounding the fundamental domain ($\chi_0=0$)
instead of  the electric one.

\begin{itemize}
  
\item $\bm{\sigma_{*,1}, \sigma \rightarrow i\infty}$: for the purely magnetic solution $\sigma=i\Theta^{4/z} \rightarrow i\infty$ at large $T$ and it is known (see appendix \S\ref{app:modular-forms}) that  $f\approx -\frac{e^{2\pi\Theta^{4/z}}}{256}$,
$\bar f'\approx  2\pi i \bar f$, so
\[ \beta \approx 2\pi i(-1)^{m+n+1} C \Theta^{8/z} \frac{e^{2\pi (m+n+1)\Theta^{4/z}}}{(256)^{m+n+1}}.\]
For $\beta$ to be analytic as $\Theta \rightarrow \infty$, $n$ and $m$ must be restricted to ensure that
\hbox{$m + n +1 < 0$}, but there are three fixed points in the fundamental domain and, as before, it will not
be possible have $\beta$ analytic at all three so $\beta$ will
not be constrained to be analytic as $T\rightarrow \infty$.
The least extreme case is the marginal one, $m+n+1=0$, for which 
\[ \xi \frac{d \Theta^{4/z}}{d \xi} = 2 \pi C \Theta^{8/z}
\quad \Rightarrow \quad \xi \approx  \xi_0 \exp\left(-\frac{1}{2 \pi C \Theta^{4/z}}\right)\]
and the coherence length increase with increasing $T$ if $C>0$ tending to a constant as $\Theta\rightarrow \infty$.

\item $\bm {\sigma \rightarrow 0}$: in the opposite limit, for  $\Theta \rightarrow 0$, $f$ vanishes as $f\approx -16 e^{-\pi/\Theta^{4/z}}$ and
$\bar f'\approx -\frac{16 \pi i} {\Theta^{8/z}} e^{-\pi/\Theta^{4/z}}$ so
\[ \beta \approx  (-1)^{m+n+1} 4^{(2m - n +2)} i  C \pi  e^{-\pi(m+1)/\Theta^{4/z}}. \]
Only $m=-1$  gives analytic behaviour for $\beta$ and, for $n=0$,   
\[\xi \frac{d \Theta^{4/z}}{d \xi} \approx  \frac{ (-1)^n \pi  C}{4^n} \quad
  \Rightarrow \quad \xi \approx \tilde \xi_0 \exp\left(\frac{(-1)^n 2^{2 n}\Theta^{4/z}}{\pi C} \right). \]

\item $\bm{\sigma_*=\frac{1+i}{2}}$: lastly there are the new fixed points at $\sigma_*=\frac{1+i}{2}$ and its images.
For $\sigma=\sigma_* + \epsilon$,
\[f\approx\frac{1}{4} - \frac{\Gamma(1/4)^8}{64 \pi^4} \epsilon^2\]
and the metric contributes a factor of $\frac 1 4 $. With $m=-1$
\[ \beta \approx  (-1)^n 2 C  \left\{\frac{\Gamma(1/4)^8}{64 \pi^4}\right\}^{n+1} \bar \epsilon^{2 n+1}.\]
Analyticity requires $n\ge 0$ and \hbox{$m+n+1=0$}
imposes  $n=0$, giving scaling behaviour
\beq \beta \approx  -\frac{\Gamma(1/4)^8}{32 \pi^4} C \bar \epsilon.\label{eq:Gamma02*}\eeq
In contrast to the $\Gamma(1)$ case the temperature dependence of $\bar \epsilon$ near $\sigma_*$
for $\Gamma_0(2)$ can be determined by going back to the full $\Gamma(1)$ configuration
in \S\ref{sec:Gamma1} and mapping its template flow ({\it i.e.} the positive imaginary axis
together with its fixed point at $\sigma_{*,2}=i$) to the semi-circle
that is the lower bound of the fundamental domain of $\Gamma_0(2)$ in figure \ref{fig:Gamma02-naive-flow}
({\it i.e.} the semi-circular arc  of radius one-half spanning
the two points $\sigma=0$ and $\sigma=1$).
This is achieved by using the $\Gamma(1)$ transformation $\bpm 0 & 1 \\ -1 &\  1 \epm$, which
is not in $\Gamma_0(2)$,  and sends $\sigma_{*,2}$ to $\sigma_*=\frac{1+i}{2}$.
The semi-circle can then be mapped around the upper half-plane using $\Gamma_0(2)$.

To apply equation (\ref{eq:Gamma02*}) in this topology 
observe that $\bpm 0 & 1 \\ -1 &\  1 \epm$ sends 
\[ i\Theta^{4/z} \rightarrow \frac{1 + i \Theta^{4/z}}{1+\Theta^{8/z}}
  \qquad \mbox{and} \qquad
  i\Theta^{-4/z} \rightarrow \frac{1+i\Theta^{-4/z}}{1+\Theta^{-8/z}}\]
with $0\le \Theta \le1 $.
Near the critical point $\sigma_{*,2}=i$, where $\Theta=1-{\tt t}$, the image is
\[\left(\frac{1+i}{2}\right) \pm  \frac{2 {\tt t}}{z} \]
with  positive ${\tt t}$ just below $\sigma_{*,2}$ in the $\Gamma(1)$ template
mapped to the right-hand half of the semi-circle ($+$ sign)
and positive ${\tt t}$ just above $\sigma_{*,2}$ mapped to the left-hand half ($-$ sign),
with $\bar \epsilon = \pm \frac{2 {\tt t}}{z}$ when ${\tt t}$ is small.
In either case the sign in (\ref{eq:Gamma02*})
makes $\frac{1+i}{2}$ a repulsive fixed point if $C$ is positive and
\[\xi \sim {\tt t}^{-\nu'} \] 
with critical exponent
\[ \nu' = \frac{32 \pi^4}{C \{\Gamma(1/4)\}^8}.\]

\end{itemize}
In summary the $\Gamma_0(2)$ flow with $m=-1$ and $n=0$ has $\beta$-function
\beq \beta = C \{Im(\sigma)\}^2 \frac{d \ln |\bar f|}{d \bar \sigma}\label{eq:Gamma02-beta-function}\eeq
with $C>0$ and the fixed point behaviour of the coherence length is
\beq \xi \sim
  \begin{cases}
    \xi_0 \exp \left(-\frac{1}{2 \pi C \Theta^{4/z}}\right), &   \Theta \rightarrow \infty, \ \sigma \rightarrow i\infty;\\
    \tilde \xi_0 \exp \left(\frac{\Theta^{4/z}}{\pi C}\right), &  \Theta \rightarrow 0, \ \sigma \rightarrow 0; \\
    A {\tt t}^{-\nu\,'}, &  \Theta \rightarrow 1, \ \sigma \rightarrow \frac{1+i}{2},\ \mbox{with} \ \nu' =\frac{32 \pi^4}{C \bigl\{\Gamma(\frac{1}{4})\bigr\}^8 }.
  \end{cases}\label{eq:Gamma02-fixed-point-asymptotics}\eeq

\begin{figure}[h]
\centerline{\includegraphics[width=10cm]{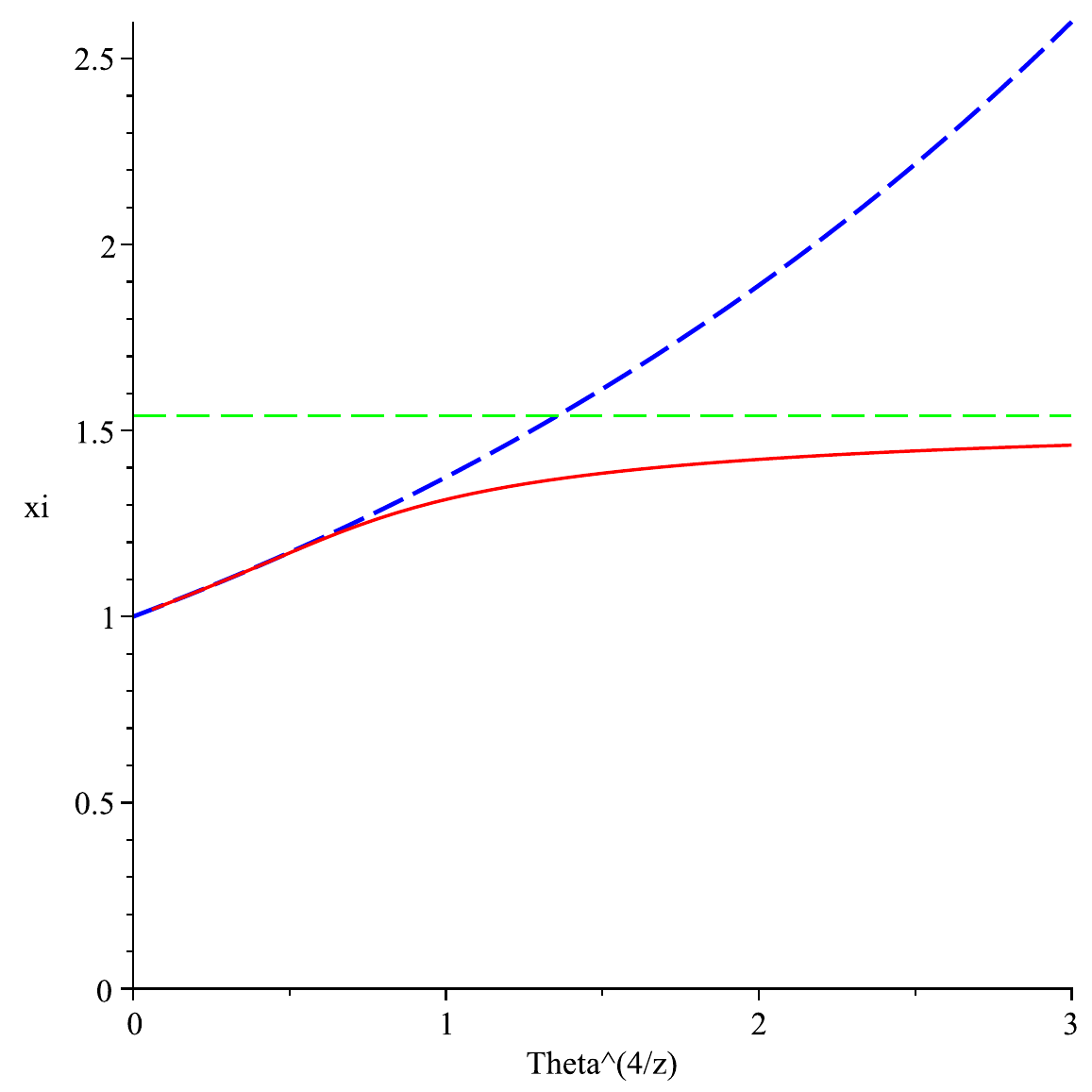}}
\medskip
\begin{caption}  { \small $\xi$ along the imaginary axis $\sigma=i \Theta^{4/z}$ for
    the purely magnetic solution as a function of $\Theta^{4/z}$. $\Gamma_0(2)$ symmetry
    determines the behaviour near $\Theta=0$, via (\ref{eq:Gamma02-fixed-point-asymptotics}),
    and (\ref{eq:Gamma02-beta-function}) is numerically integrated to determine $\xi\bigl(\Theta^{4/z}\bigr)$.
    The red curve is the numerical result for the specific choice $\tilde \xi_0=C=1$. The dashed blue curve is the asymptotic
    form $\xi = e^{\Theta^{4/z}/\pi}$ near $\Theta=0$ and the dashed green curve is the asymptotic value for
    $\tilde \xi_0=C=1$, which numerically is $\xi_0 \approx 1.540$. Again this assumes that the classical solution gives the correct temperature dependence along the imaginary axis.
     }
\label{fig:xi_v_theta}\end{caption} 
\end{figure} 

\begin{figure}[h]
\centerline{\includegraphics[width=14cm]{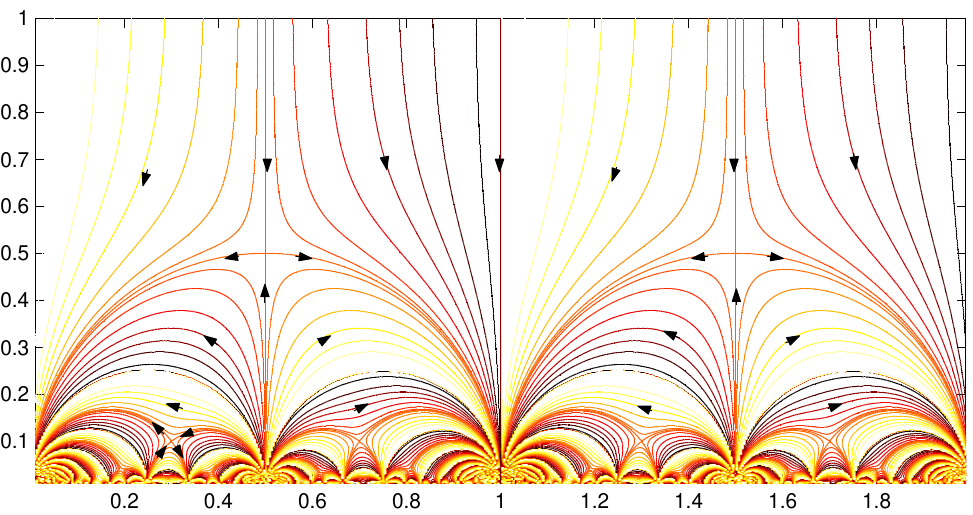}}
\medskip
\begin{caption}  { \small Temperature flow of the conductivity on the event horizon
    for $\Gamma_0(2)$ symmetry, assuming it is generated by varying $|f|$, keeping $\arg(f)$ fixed.
    The diagram is obtained simply by plotting lines of constant $\arg(f)$ as in \cite{Crossover}.}
\label{fig:Gamma02-full-flow}\end{caption} 
\end{figure}

Extending the temperature into the whole fundamental domain by assuming that $\Theta(f,\bar f)$ is
a real modular invariant function that increases monotonically as $|f|$ increases
the flow can be obtained by plotting lines of constant $\arg(f)$ in the conductivity plane.
The result is shown in figure \ref{fig:Gamma02-full-flow} which is taken from the review \cite{BD-SIGMA} and has the same topology as the $\Gamma_0(2)$ flow originally suggested in \cite{Lutken1,Lutken2}.  Flow lines in the conductivity plane, in terms of electron coherence length,
were suggested for the integer QHE in \cite{Khmelnitskii} and the hierarchical structure of modular symmetry extends this to the fractional effect.

Note that it is not being assumed that $\Theta$ is independent of $\arg(f)$, only that the temperature flow is obtained by varying $|f|$, keeping $\arg(f)$ fixed. For $\sigma =i\Theta^{4/z}$, $f$ is real and negative, so $\arg(f)=\pi$, while on the semi-circle passing through $\sigma_*$, $f$ is real and positive, so $\arg(f)=0$.  The anti-holomorphic $\beta$-function in (\ref{eq:Gamma02-beta-function}) was first proposed in \cite{L+R:anti-holomorphic1,L+R:anti-holomorphic2,Lutken3}.

Figure \ref{fig:Gamma02-full-flow} should be compared to the experimental plots in \cite{Murzin-integer}
and \cite{Murzin-fractional} (similar experimental plots have been obtained by other groups \cite{Taiwan-group}). The agreement is remarkable.

\section{Discussion\label{sec:Discussion}}

There are two distinct aspects to the discussion presented here: the AdS/CMT paradigm and gradient flow.
They are woven together in the analysis but are a priori independent concepts.

The idea of gradient flow $\beta$-functions for the QHE is motivated
by the $c$-theorem in 2-dimensions and was first used in the context of modular symmetry more than 20
years ago \cite{Burgess+Lutken1,Burgess+Lutken2}. In that work the function $f'/f$  was considered\footnote{Denoted by $E^T_2$ in \cite{Burgess+Lutken1,Burgess+Lutken2}.} but rejected it because it did not match perturbation theory for large values of the Ohmic conductivity.\footnote{Holomorphic  modular forms of weight -2 for $\Gamma_0(2)$ were used in \cite{Crossover} in a discussion of how the conductivity changes between quantum Hall plateaux as the magnetic field is varied at fixed temperature, where the same problem was noted.
This also gave a pole at $\sigma=\frac{1+i}{2}$ and a mechanism was proposed to tame the pole and obtain smooth crossovers between plateaux at finite $T$, well away from the perturbative limit.  This was based on a holomorphic rather than an anti-holomorphic ansatz.} In the context of the AdS/CMT correspondence that is not
a reason to reject it as the perturbative limit of the boundary theory is not accessible from the classical bulk theory, analysing the perturbative limit would require understanding quantum gravity effects in the bulk.

As mentioned in the introduction a number of authors have considered describing
the QHE within the framework of AdS/CMT, including in particular \cite{GIKPTW} where the Gibbons-Rasheed action was used near the horizon to analyse the conductivity associated with a dyonic solution in the bulk and \cite{Bak+Rey} where the Dirac quantisation condition in the bulk was used to argue for fractional filling factors. The new ingredients here are the observation that the infra-red DC conductivity at $u\rightarrow 1$ is identified with the value of the dilaxion on the horizon and that this is consistent with gradient flow.

The assumption that $\Theta$ is a monotonic function of $|f|$ and that the flow
lines are given by varying $|f|$ keeping $\arg(f)$ fixed needs some discussion (we shall discuss $\Gamma_0(2)$ here, as that is more relevant to experiment, a similar discussion can be given for $\Gamma(1)$ with $f$ replaced by $J$).
$\Theta$ is a monotonic function of $|f|$ on the positive imaginary axis, where $\arg(f)=\pi$, and extending this into the interior of the fundamental domain requires turning on a non-zero $\chi_0$ and changing $\arg(f)$, so that $\arg(f)$ is correlated in some way with $\chi_0$ in the classical solution.
Experimentally \cite{Murzin-integer}-\cite{Taiwan-group} this can be achieved in the QHE by varying the magnetic field away from its critical value $B_*$.
In the composite fermion picture $B_*$ corresponds to the situation when the statistical
gauge field exactly cancels the applied field, the 2DEG is a composite fermi liquid \cite{Bak+Rey}.
In our scenario $B_*$ corresponds to $\lambda \chi_0 = 1/2$ and deviations
from $B_*$ arise not from varying the dyon magnetic charge, which relates to a global $U(1)$ on the boundary, but from varying $\chi_0$ (in \cite{Bak+Rey} $\chi_0$ is set to zero).  Exactly
how $\chi_0$ might be related to the deviation of the external field from its critical value
would depend on the details of the underlying matter and an investigation of this would require
numerical analysis of an underlying fermionic matter action.
But the basic idea that varying $T$ and $B$ independently is equivalent to varying
$\x$ and $\chi_0$, or $|f|$ and $\arg(f)$, is not obviously inconsistent and seems at least plausible.

Lastly we comment on possible sub-groups of $\Gamma(1)$. Two scenarios have been presented involving $\Gamma(1)$ and $\Gamma_0(2)$ flow with very different
topologies, but there are other possibilities.
For example $\Gamma(2)$, generated by ${\bm S}{\bm T}^2{\bm S}$ and ${\bm T}^2$, was studied in \cite{Georgelin} in the
context of the QHE and is relevant for spin-degenerate quantum Hall systems \cite{spin-splitting}.
Another level 2 sub-group, $\Gamma_\theta$ (see appendix \ref{app:modular-forms}), was suggested as being relevant for bosonic charge carriers
in a superconductor in a magnetic field in \cite{1917}.  The possibility of ${\bf T}$ versus ${\bf T}^2$ was discussed in \cite{Witten-Sl2Z} where it was related to whether or not the Euclidean version of the bulk theory is formulated on  a 4-manifold admitting a spin structure. 
There is nothing in the classical solutions presented here
that picks out any preferred level 2 sub-group associated with any particular kind of matter, the dilaton
is not charged and merely plays the r\^ole of an effective  background electric susceptibility after any
matter fields have been integrated out. Presumably a more detailed bulk model, included matter, is needed to pick
out a specific level 2 sub-group.  For example fermionic matter in the bulk would be expected to  give $\Gamma_0(2)$ and bosonic matter $\Gamma_\theta$.  The full group $\Gamma(1)$ might require supersymmetric matter, though a
phase diagram with $\Gamma(1)$ symmetry was proposed for a
2-dimensional Abelian lattice model in \cite{Cardy} (to the author's knowledge this was actually the first suggestion of $Sl(2,{\bm Z})$ duality transformations between different phases)
but this is left as a subject for future investigation.

\appendix

\section{Conventions \label{app:conventions}}

The conventions used in the text for the computation of the conductivity
in appendix \ref{app:LRT}
are collected here for convenient reference.

Our signature is $(-,+,+,+)$ with line element
\beq
d s =\frac{1}{u^2} \left( - f^2 d \tilde t\,^2 + \frac{l^2}{h^2} d u^2 +  d x^2 + d y^2 \right),  
\eeq  
where $u=\frac{r_h}{r}$ and $f$ has a single zero at $u=1$,
with asymptotic infinity at $u\rightarrow 0$.
The magnetic field tensor $F_{\mu\nu}$ decomposes into electric and magnetic fields as
\[ E_\ca = -F_{\tilde t \ca}, \qquad B^\ca = - \widetilde F^{\tilde t \ca}
=-\frac{1}{2}\frac{\epsilon^{\tilde t \ca \cb \cc}}{\sqrt{-g} } F_{\cb \cc} \]
with $\ca,\cb,\cc=u,x,y$ and $\epsilon_{\tilde t u x y}=-\epsilon^{\tilde t u x y} =1$.
Under a perturbations homogeneous in the transverse directions,
the transverse field variations are 
\[ \delta E_\ca = -\delta F_{\tilde t \ca}, \qquad
  \delta B^\ca = - \delta \widetilde F^{\tilde t \ca}
  = -\frac{u^4 h}{l f} \epsilon^{\ca \cb} (\delta F_{u \cb})
= -\frac{u^4 h}{l f} \epsilon^{\ca \cb} (\delta A_\cb')
\]
with $\epsilon^{x y}=-\epsilon^{y x}=1$ and $\ca,\cb$ restricted to $x,y$.

Conductivities are calculated in a local inertial frame and we need these
expressions in an orthonormal basis.  With transverse orthonormal
indices $i,j=2,3$ the orthonormal components of ${\bf E}$ and ${\bf B}$ are
\[ \delta E_i = \frac{u^2}{f} \delta_i{}^\ca (\delta E_\ca - \tm \epsilon_\ca{}^\cb \delta G_\cb), \qquad
  \delta B^i = \frac{f}{u^2} \delta^i_\ca(\delta B^\ca) 
= -\frac{u^2 h}{l} \epsilon^{\ca \cb} (\delta A_\cb'),
\]
where the metric variation is
$\delta g_{\tilde t \ca} = \frac{1}{u^2}\delta G_\ca$ and $\tm$ is related to the magnetic charge $m$
in equation (\ref{eq:F-u}) by $\tm = \bigl( \frac{l}{r_h} \bigr)^2 m$.
It is convenient to combine these into the complex fields
\[ \E_\pm = \frac{f}{u^2}(\delta E_2 \pm i \delta E_3), \qquad
  \B_\pm = \frac{f}{u^2}(\delta B_2 \pm i \delta B_3)
  = \frac{f^2}{u^4}(\delta B^x \pm i \delta B^y).
\]

The Hodge star used in (\ref{eq:*G}) is
\[*1 = \sqrt{-g}\,d^4 x= \frac{l}{u^{z+3}}
(d \tilde t \wedge d u \wedge d x \wedge d y),\]
thus $*G=\frac{ \sqrt{-g}}{2}  \epsilon_{\mu\nu\rho\sigma}  G^{\mu\nu}d x^\rho \wedge d x^\sigma = \frac{1}{2}{\widetilde G}_{\rho\sigma} d x^\rho \wedge d x^\sigma$.

\section{Conductivities in linear response theory\label{app:LRT}}

Conductivities in the boundary theory can be determined using the techniques in \cite{H-K-1} and
\cite{H-H}.  The idea is to make a perturbation of the fields that is independent of 
the transverse co-ordinates $x$ and $y$ and demand that this is a solution of the 
bulk equations of
motion at second order in the perturbation. This generates differential equations that the
perturbations must satisfy which can be used to determine response functions
for the boundary theory, in particular the conductivities.

Let
\[ \delta g_{\tilde t x}= \frac{\delta G_x(\tilde t,u)}{u^2} \qquad
  \delta g_{\tilde t y}= \frac{\delta G_y(\tilde t,u)}{u^2}\]
be perturbations of the bulk metric
\[  d s^2=\frac{1}{u^2}\left( - f^2(u) d \tilde t\,^2 + \frac{l^2 d u^2}{ h^2(u)} + d x^2 + d y^2 \right).\]
These are not general perturbations but are chosen to be independent of $x$ and $y$
so they are homogeneous in the transverse directions, which is sufficient for our needs. We shall also assume oscillatory time dependence and
set
\[ G_x(\tilde t,u) =e^{-i\widetilde \omega \tilde t} \delta \widetilde G_x(u),\qquad 
   G_y(\tilde t,u) =e^{-i\widetilde \omega \tilde t} \delta \widetilde G_y(u).\] 
For the explicit dyon solution in \S\ref{sec:Dyon} the functions $f(u)$ and $ h(u)$ are
\[ f^2(u)=\frac{(1-u^{z+2})}{u^{2(z-1)}}, \qquad  
 h^2(u)= 1-u^{z+2}\]
but the analysis will be kept more general and these explicit forms only used at the
end.
Perturbations of the dilaton and axion fields take a similar form
\beq \delta \phi(\tilde t,u) = e^{-i\widetilde \omega \tilde t} \delta \tilde \phi(u),
\qquad  
\delta \chi(\tilde t,u) = e^{-i\widetilde \omega \tilde t} \delta \tilde \chi(u),\label{eq:delta-dilaxion}\eeq
though demanding that the perturbations satisfy the linearised equations of motion
forces $\delta \tilde \phi(u)=\delta \tilde \chi(u)=0$ at this order.

A similar perturbation of a dyonic configuration, with Maxwell 2-form\footnote{In the solution
  presented in \S\ref{sec:Dyon}, equation (\ref {eq:F-u}),
\[\tq = \left(\frac{r_h}{l}\right)^2 q, \qquad \tm = \left(\frac{l}{r_h}\right)^2 m .\]}
\[F = d A = \frac{\tq\, l f(u) }{u^4  h(u)}d \tilde t \wedge d u + \tm d x \wedge d y,\]
is 
\beq\delta A 
= \delta A_\ca (\tilde t,u) d x^\ca =e^{-i\widetilde \omega \tilde t}\Bigl(\wa_x(u) d x + \wa_y(u) d y\Bigr).
\label{eq:delta-A}\eeq
This generates a transverse electric field
\beq\delta E_\ca = -\delta F_{\tilde t \ca}=i\widetilde \omega \delta A_\ca\label{eq:delta-E}\eeq
and a transverse magnetic field
\beq \delta B^\ca = \frac{\epsilon^{\tilde t u \ca \cb }}{\sqrt{-g}} \delta F_{u \cb } 
=-\frac{u^4  h}{l f}\epsilon^{\ca \cb} \delta A'_\cb ,\label{eq:delta-B}\eeq
where where $\ca$ and $\cb$ label $x$ and $y$, $\epsilon^{\tilde t u x y}=-1$, $\epsilon^{x y}=+1$ and  $'$ denotes differentiation with respect to $u$.
For static fields
 \beq  \delta A_\ca(\tilde t,u) = (\delta E_\ca^0) \tilde t + \delta \widetilde A_\ca(u)\eeq
with $\delta E^0_\ca$ constants. 

In an orthonormal basis the electric and magnetic field variations are 
         \begin{eqnarray}
           \delta E_2 =\frac{u^2}{f}(\delta E_x - \tm \delta G_y),\qquad
           \delta E_3 &=&\frac{u^2}{f}(\delta E_y + \tm \delta G_x),\label{eq:delta-E-ortho}\\
           \delta B^2 =-\frac{u^2  h}{l} \delta A_y'=\frac{f}{u^2} \delta B^x,\qquad
           \delta B^3 &=& \frac{u^2  h}{l} \delta A_x'=\frac{f}{u^2} \delta B^y.\label{eq:delta-B-ortho}
           \end{eqnarray}

These perturbations will induce a transverse current
\[\delta J^i = \sigma^{i j} \delta E_j \]
with $\sigma^{i j}$ the transverse conductivity tensor ($i,j=2.3$ are orthonormal indices). 
In a co-ordinate basis  
\[ \bpm \delta J^x \\\delta J^y \epm = u\bpm \delta J^2\\\delta J^3\epm,
  \qquad \bpm \sigma^{x x} & \sigma^{x y} \\ \sigma^{y x} & \sigma^{y y} \epm 
= u^2 \bpm \sigma^{2 2} & \sigma^{2 3} \\ \sigma^{3 2} & \sigma^{3 3} \epm\]
and 
\[\delta J^\ca = \sigma^{\ca \cb} \E_\cb \]
where
\beq \E_\ca = \delta E_\ca -\tm \epsilon_\ca{}^\cb \delta G_\cb.\eeq
With the complex combinations
\beq
 \E_\pm = \E_x \pm i \E_y,\qquad 
\delta J_\pm  = \delta J^x \pm i \delta J^y, \qquad    \sigma_\pm = \sigma^{x y} \pm i \sigma^{x x}
\eeq
this is \cite{H-H}
\beq \sigma_\pm = \pm i \frac{\delta J_\pm}{\E_\pm}.\eeq

In AdS/CMT correlations functions are derived from the boundary action.
With any fixed value of  $\delta A_\ca|_{u=1}$ at the event horizon we can cut
the bulk theory off at a finite value of $u<1$ to find that the variation of the action is
\begin{align*} \delta S[\delta A(u),\delta G(u)] &=  \int d \tilde t\, d x\, d y 
\left[ \delta A_\ca \left( -\frac{1}{2} e^{-\lambda \phi}\delta F^{u \ca} \sqrt{-g}
+ \frac{\lambda \chi}{2} \epsilon^{u \ca \tilde t \cb}  \delta F_{\tilde t \cb}\right) \right]_u^1\nonumber \\
&=\frac{1}{2}\int d \tilde t\, d x\, d y
\left.\Bigl\{ \delta A_\ca \left(e^{-\lambda \phi}\delta F^{u \ca} \sqrt{-g}
 -\lambda \chi \epsilon^{u \ca \tilde t \cb}  \delta F_{\tilde t \cb}\right) \Bigr\}\right|_u  + const.\label{eq:boundary-term}
\end{align*}
(the Einstein action gives no contribution at $O(\delta G_\ca^2)$\,).
The current is then
\begin{align}
 \delta J^\ca(u) & = \frac{\delta S}{\delta A_\ca}
=e^{-\lambda \phi} \delta F^{u \ca} \sqrt{-g} - \lambda \chi \epsilon^{\tilde t u \ca \cb} \delta F_{\tilde t \cb}\\
& =\frac{e^{-\lambda \phi}}{u^4}\delta^{\ca\cb}
\left\{ f^2 \epsilon_{\cb \cc} (\delta B^\cc) - \tq \, (\delta G_\cb) \right\}
- \lambda \chi \epsilon^{\ca \cb} (\delta E_\cb) 
\end{align}
where we have used 
\[ \delta F^{u \ca} = \delta^{\ca \cb} 
\left( \frac{u^4 h^4}{l^2} \delta A_\cb -\frac{\tq\, h}{ l f}   \delta G_\ca\right),
\qquad \sqrt{-g}=\frac{l f}{u^4 h},\] 
and (\ref{eq:delta-B}). 
This can be re-expressed as
\[
\delta J^\ca = e^{-\lambda \phi}\delta^{\ca\cb}
 \epsilon_{\cb \cc} (\B^\cc) -  \lambda \chi \epsilon^{\ca \cb}\E_\cb
+\delta^{\ca \cb}  \left( \lambda \tm \chi -\frac{\tq e^{-\lambda \phi }}{u^4}\right) \delta G_\cb\]
where, following  \cite{H-H}, we define
\[ \B^\ca = \frac{f^2}{u^4} \delta B^\ca= \frac{f}{u^2} \delta^\ca_i (\delta B^i)\]
in analogy with
\[ \E_\ca = \frac{f}{u^2} \delta_\ca^i (\delta E_i).
\]

From these the transverse conductivity, ignoring the back-reaction on the metric, is
\beq 
\sigma^{\ca \cb} 
= \left.\frac{\delta J^\ca}{\E_\cb}\right|_{\delta G_\pm=0}
=e^{-\lambda \phi}  \left(\frac{\epsilon^\ca{}_\cc \B^\cc}{\E_\cb}\right)
- \lambda \chi \epsilon^{\ca \cb} \eeq
or
\beq  \sigma_\pm =e^{-\lambda \phi} \frac{\B_\pm}{\E_\pm}-\lambda \chi,
\label{eqa:sigma_pm}\eeq
which is central to the analysis in the text.

We note in passing that
\beq \left.\frac{\delta J^\ca}{\delta G_\cb}\right|_{\E_\pm=0}=\left(\lambda \tm \chi -\frac{\tq e^{-\lambda \phi}}{u^4}\right) \delta^{\ca \cb} =\frac{\tm \Sla}{\Slc}\delta^{\ca \cb}= \frac{ Q_e}{\Areah}  \delta^{\ca \cb},\eeq
where the final two equalities are specifically for the dyon solution (\ref{eq:phi-chi})
with electric charge  (\ref{eq:Q_e}).  $\frac{Q_e}{\Areah}$ is of course the charge density
at the event horizon and $u^2\frac{\delta J^\ca}{\delta G_\cb}=< J^\ca T^{\tilde t \cb}>$ is piezoelectric tensor for the deformation $\delta g_{\tilde t \cb}$.

\subsection{RG equation for the conductivities
\label{app:Conductivity-RG}}

To obtain more detailed information about  the conductivity we need a relation between $\B^\ca$ and $\E_\ca$ 
and this comes from requiring that these variations are solutions of the linearised equations of motion.
These are Einstein's equations
\begin{align}
\frac{f h}{4 \kappa^2 l^2}
\left(\frac{h}{f}\frac{ \delta G_x'}{u^2} \right)'
+e^{-\lambda\phi}\left(-\frac{f h}{l}\frac{\tq}{u^4} \delta A_x' -  \tm (i \widetilde \omega \delta A_y + \tm \delta G_x)\right)&=&0,\label{eq:EV1}\\
\frac{f h}{4 \kappa^2 l^2}
\left(\frac{h}{f}\frac{ \delta G_y'}{u^2} \right)'
+e^{-\lambda\phi}\left(-\frac{f h}{l}\frac{\tq}{u^4} \delta A_y' + \tm (i \widetilde \omega \delta A_x - \tm \delta G_y)\right)&=&0,\label{eq:EV2}\\
\frac{i\widetilde \omega }{4 \kappa^2}
\left(\frac{ \delta G_x'}{u^2} \right)
+e^{-\lambda\phi}\left(-\frac{ l f}{h}\frac{\tq}{u^4}(i \widetilde \omega \delta A_x - \tm  \delta G_y) - \tm f^2 \delta A_y'\right)&=&0,\label{eq:EV3}\\
\frac{i\widetilde \omega }{4 \kappa^2}
\left(\frac{ \delta G_y'}{u^2} \right)
+e^{-\lambda\phi}\left(-\frac{l f}{h}\frac{\tq}{u^4} (i \widetilde \omega \delta A_y + \tm \delta G_x) + \tm f^2 \delta A_x'\right)&=&0,\label{eq:EV4}
\end{align}
and Maxwell's equation
\begin{align}
  \partial_\mu\bigl(\sqrt{-g}\, e^{-\lambda\phi} \delta F^{\mu\nu}\bigr)-\frac{\lambda}{2} \epsilon^{\mu\nu\sigma\lambda} (\partial_\mu \chi) \delta F_{\sigma\lambda}=0 \qquad   \Rightarrow  & \nonumber\\
  -\frac{i \widetilde \omega l}{f h} e^{-\lambda \phi}
  \bigl(i\widetilde \omega \delta A_\ca - \tm \epsilon_\ca{}^\cb \delta G_\cb\bigr)
  +\left(\frac {f h}{l}  e^{-\lambda \phi} (\delta A_\ca)'  - \frac{\tq}{u^4} \delta G_\ca \right)'
  & = i\widetilde \omega \lambda \chi' \epsilon_{\ca}^\cb \delta A_\cb, \label{eq:Max}
\end{align}
while the dilaton and axion equations of motion impose
\beq \delta \phi = \delta \chi =0\label{eq:delta-cphi}\eeq
(see \cite{H-K-1}, the only new ingredient here is the dilaton and axion for which a perturbation
of the form (\ref{eq:delta-dilaxion}) is
constrained to vanish by the equations of motion).

In terms of 
\bean
\E_x &=& i\widetilde \omega \delta A_x - \tm \delta G_y,\label{eq:delta-E-G2}\\ 
\E_y &=& i\widetilde \omega \delta A_y + \tm \delta G_x,\label{eq:delta-E-G3}\\ 
\B^x &=& -\frac{f h}{l} \delta A_y',\\
\B^y &=& \frac{f h}{l} \delta A_x',
\eean
equations (\ref{eq:EV1})-(\ref{eq:EV4}) can be re-cast as
\begin{eqnarray}
\frac{f h}{4 \hat \kappa^2}
\left(\frac{h}{f}\frac{ \delta G_x'}{u^2} \right)'
&=&e^{-\lambda\phi}\left(\tm\E_y +\frac{\tq}{u^4} \B^y\right),\label{eq:EV1-deltaG}\\
\frac{f h}{4 \hat \kappa^2}
\left(\frac{h}{f}\frac{ \delta G_y'}{u^2} \right)'
&=& e^{-\lambda\phi}\left(- \tm \E_x - \frac{\tq}{u^4} \B^x\right),\label{eq:EV2-deltaG}\\
\frac{i\womega }{4 \hat \kappa^2}\frac{h}{f}
\left(\frac{ \delta G_x'}{u^2} \right)
&=& e^{-\lambda\phi}\left(\frac{\tq}{u^4} \E_x - \tm  \B^x\right),\label{eq:EV3-deltaG}\\
\frac{i\womega }{4 \hat \kappa^2}\frac{h}{f}
\left(\frac{ \delta G_y'}{u^2} \right)
&=&e^{-\lambda\phi}\left(\frac{\tq}{u^4} \E_y - \tm  \B^y\right),\label{eq:EV4-deltaG}
\end{eqnarray}
with $\hat\kappa = \kappa l$ and $\womega = \widetilde \omega l$
while (\ref{eq:Max}) is
\[
   -\frac{i \womega}{f h} e^{-\lambda \phi}(\E_\pm)
  +\bigl(\epsilon_\ca{}^\cb \B_\cb - \frac{\tq}{u^4} \delta G_\ca \bigr)'
 = i\womega \lambda \chi' \epsilon_{\ca}^\cb (\E_\cb +\tm \epsilon_\cb{}^\cc \delta G_\cb).
\]

Now differentiate (\ref{eq:EV3-deltaG}) and (\ref{eq:EV4-deltaG}) and equate the result to
(\ref{eq:EV1-deltaG}) and (\ref{eq:EV2-deltaG}) giving
\beq
\left\{ e^{-\lambda \phi} \left(- \frac{\tq}{u^4} \E_\pm + \tm \B_\pm \right)\right\}' = \mp \frac{\womega e^{-\lambda\phi}}{f h}\left( \tm \E_\pm + \frac{\tq}{u^4}\B_\pm \right).
\label{eq:E'B'}\eeq 
A second equation relating $\E_\pm'$ to $\E_\pm$
and $\B_\pm$ is obtained from (\ref{eq:delta-E-G2}) and (\ref{eq:delta-E-G3}),
\beq
\delta G_x = \frac{1}{\tm} (-i\widetilde \omega \delta A_y + \E_y),\qquad
\delta G_y = \frac{1}{\tm} (i\widetilde \omega \delta A_x - \E_x),
\eeq
from which
\beq
\delta G_x' = \frac{1}{\tm} \left(\frac{i\womega}{f h} \B^x + \E_y'\right),\qquad
\delta G_y' = \frac{1}{\tm} \left(\frac{i\womega}{f h} \B^y- \E_x'\right).\label{eq:delta-G}
\eeq

Now using these to eliminate $\delta G_\ca'$ in (\ref{eq:EV3-deltaG}) and (\ref{eq:EV4-deltaG}) leads to
\beq
\frac{\womega}{4 \hat \kappa^2 u^2}\left(\mp \E_\pm' + \frac{\womega}{f h} \B_\pm \right) = e^{-\lambda \phi} \frac{f}{h} \left(-\frac{\tq \tm}{u^4} \E_\pm + \tm^2 \B_\pm \right)\label{eq:E'B}.
\eeq
Equations (\ref{eq:E'B'}) and (\ref{eq:E'B}) here are the analogues of equations (17) and (18) in 
\cite{H-H}. 

It is convenient to define
\begin{equation}\cQ=\frac{e^{-\lambda\phi} \tq}{u^4}, \qquad
\cM=e^{-\lambda\phi} \tm,\qquad
\delta {\cal G}_\pm = \delta {\cal G}_x \pm i \delta {\cal G}_y,
\label{eq:QM-def}\end{equation}
in terms of which (\ref{eq:E'B'}), (\ref{eq:delta-G}) and  (\ref{eq:E'B}) can be written as a matrix equation
\begingroup
\renewcommand*{\arraystretch}{1.5}
\begin{align}
\begin{pmatrix} -\cQ & \cM & 0 \\
\mp\frac{\womega}{4 \hat \kappa^2 u^2} & 0 &0 \\
1 & 0 & \mp i \tm \end{pmatrix} \begin{pmatrix} \E'_\pm \\ \B'_\pm \\ \delta  {\cal G}'_\pm\end{pmatrix} 
&=\begin{pmatrix} \cQ' \mp \frac{\womega}{f h}\cM & 
-\cM' \mp \frac{\womega}{f h}\cQ& 0\\   
-\frac{f \tm}{h}  \cQ & \frac{f \tm}{h}  \cM -\frac{\womega^2}{4\hat \kappa^2 u^2 f h} &0\\
-\frac{4i\hat \kappa^2 u^2 f}{\womega h} \cQ &
 \frac{4i\hat \kappa^2 u^2 f}{\womega h} \cM& 0
\end{pmatrix} 
\begin{pmatrix} \E_\pm \\ \B_\pm \\\delta {\cal G}_\pm\end{pmatrix} \nonumber \\
\Rightarrow \qquad 
 \begin{pmatrix} \E'_\pm \\ \B'_\pm \\\delta  {\cal G}'_\pm\end{pmatrix} & = 
 \begin{pmatrix} {\cal S}_{11}   & 
{\cal S}_{1 2} & 0 \\
{\cal S}_{2 1} &
{\cal S}_{2 2} & 0 \\
{\cal S}_{3 1} &
{\cal S}_{3 2} & 0
\end{pmatrix}
\begin{pmatrix} \E_\pm \\ \B_\pm \\ \delta {\cal G}_\pm \end{pmatrix} \label{eq:matrix-ODE}
\end{align}
\endgroup
with
\begin{align}
{\cal S}_{1 1} &= \pm\frac{4\hat \kappa^2 u^2}{\womega} \frac{f \tm}{h} \cQ   \nonumber \\
{\cal S}_{1 2} &= \mp\frac{4\hat \kappa^2 u^2}{\womega} \frac{f \tm}{h} \cM \pm\frac{\womega}{f h}  \nonumber \\
{\cal S}_{2 1}&=\pm\frac{4\hat \kappa^2 u^2}{\womega} \frac{f \tm}{h} \frac{\cQ^2}\cM +\frac{\cQ'}\cM \mp\frac{\womega}{f h} \nonumber \\
{\cal S}_{2 2}&=\mp\frac{4\hat \kappa^2 u^2}{\womega} \frac{f \tm}{h} \cQ-\frac{\cM'}\cM\label{eq:A_ij}\\
{\cal S}_{3 1} &= -\frac{4i\hat \kappa^2 u^2 f}{\womega h} \cQ \nonumber \\
{\cal S}_{3 2} &= \frac{4i\hat \kappa^2 u^2 f}{\womega h} \cM. \nonumber 
\end{align}

From these we get
\begin{align} \left( \frac{\B_\pm}{\E_\pm}\right)' & = \\
& \kern -50pt   \pm\frac{4\hat \kappa^2 u^2}{\womega} \frac{f}{h}\frac{\tm}\cM\left\{ -\cQ + \cM\left( \frac{\B_\pm}{\E_\pm}\right)\right\}^2
\mp \frac{\womega}{f h}\left\{ 1 + \left(\frac{\B_\pm}{\E_\pm}\right)^2\right\}
-\frac{\cM'}\cM\left( \frac{\B_\pm}{\E_\pm}\right)+\frac{\cQ'}{\cM}.\nonumber 
\end{align}
A renormalisation group equation for the conductivity tensor follows from this \cite{BD-I}.
Again with
\[ \Sigma_\pm = e^{-\lambda \phi} \frac{\delta {\cal B}_\pm}{\delta {\cal E}_\pm}
\]
using (\ref{eq:QM-def}) gives
\[
\left(\Sigma_\pm -\frac{\tq e^{-\lambda \phi}}{\tm u^4}\right)'
=\pm\frac{4\hat \kappa^2 u^2}{\womega} \frac{f}{h}
\left( \tm \Sigma_\pm  -\frac{\tq e^{-\lambda \phi}} {u^4} \right)^2
\mp \frac{\womega e^{-\lambda\phi}}{f h}
\left(1 + e^{2 \lambda \phi} \Sigma_\pm^2\right) \]
Putting the dyon solution 
\[ f(u) = \frac{\sqrt{1-u^{z+2}}}{u^{z-1}},\qquad h(u) = \sqrt{1-u^{z+2}},\qquad
e^{-\lambda\phi} = \frac{u^4}{\Sld^2 + \Slc^2  \x^2 u^8}\]
into this results in
\beq\pm\frac{\womega}{\tm}\left(\tm\Sigma_\pm -\frac{\tq e^{-\lambda \phi}}{u^4}\right)'
=\frac{4\hat \kappa^2}{u^{z-3}} \left(\tm\Sigma_\pm - \frac{\tq e^{-\lambda \phi}}{u^4}\right)^2
- \frac{\womega^2 e^{-\lambda\phi} u^{z-1}}{(1-u^{z+2})}\bigl( 1 + e^{2\lambda\phi}\Sigma_\pm^2\bigr).\label{eq:Sigma-dE-dB-II}
\eeq

This will lead immediately to the radial RG for the conductivity but before finally deriving that 
we pause to note that $\delta {\cal G}_\pm$ has disappeared from the analysis, because
it does not contribute to Einstein's equations.
It  does appear in Maxwell's equation
\[ \pm\frac{\womega}{f h} e^{-\lambda \phi} \E_\pm
  + \left\{ e^{-\lambda \phi} \B_\pm \mp i\frac{\tq}{u^4} \delta {\cal G}_\pm \right\}'
    = \lambda \chi'(\E_\pm \mp i \tm \delta {\cal G}_\pm)
  \]
  and if we try to use this, together with (\ref{eq:matrix-ODE}),
  to determine $\delta {\cal  G}_\pm$ we  get
  \[ \tm ( {\cal Q}' - \tm \lambda \chi')\delta {\cal G}_\pm = \mp i({\cal Q}' \tm - \lambda \chi' )\E_\pm,\]
  but this does not mean $\tm\delta {\cal G}_\pm = \mp i\E_\pm$ because
  in the dyonic solution (\ref{eq:phi-chi})
  \[ {\cal Q} - \tm\lambda \chi  = \frac{\tq \Sla}{\Slc} \]
  is constant in the dyon background.
  Although $\delta {\cal G}'_\pm$ is determined by Einstein's equations, $\delta {\cal G}_\pm$
  itself is not and it can be changed by any constant without affecting the analysis.

  Using the dyonic solution in \S \ref{sec:Dyon} we can now write (\ref{eq:Sigma-dE-dB-II}) as
  an equation for $\sigma_\pm$.  From (\ref{eqa:sigma_pm}) 
\[\sigma_\pm = \Sigma_\pm -\lambda \chi \] 
and
\begin{align*}
  \Sigma_\pm-\frac{\tq}{\tm}\frac{e^{-\lambda\phi}}{u^4}
  &=\Sigma_\pm-\frac{\tq}{\tm}\frac{ \x}{(\tSld^2 + \Slc^2   \x^2u^8)} \\
&= \sigma_\pm - \frac{\tq}{\tm}\frac{ \x}{(\tSld^2 + \Slc^2  \x^2 u^8)} + \lambda \chi \\
&= \sigma_\pm + \frac{\tSld}{\Slc }\frac{1}{(\tSld^2 + \Slc^2  \x^2 u^8)} + \frac{\tSlb\tSld + \Sla \Slc  \x^2 u^8}{\tSld^2 + \Slc  \x^2 u^8} \\
&= \sigma_\pm + \frac{\Sla  }{\Slc } = \sigma_\pm + \frac{\ia }{\ic }.\\
\end{align*}
Also
\[ 1 + e^{2 \lambda \phi} \Sigma^2_\pm = e^{2 \lambda \phi} \bigl(
e^{-2 \lambda \phi} + (\sigma_\pm + \lambda \chi)^2 \bigr)
= e^{2 \lambda \phi}(\sigma_\pm + \da)(\sigma_\pm + \overline \da).\]
Using (\ref{eq:qm-delta-gamma-I}) to express $\tm =\bigl(\frac{l}{r_h}\bigr)^2 m$ in terms of $c$, (\ref{eq:Sigma-dE-dB-II}) finally results in the following first order differential equation for $\sigma_\pm$,
\begin{align}
  \pm 
  \womega (1-u^{z+2}) u^{z-3}\sigma_\pm'  = & 4  \x (z-1)(z+2) (1-u^{z+2})
\left(\Slc \sigma_\pm + {\Sla}\right)^2  \label{aeq:sigma_pm'} \\
& \hskip 20pt - \womega^2 u^{2z-8}(\tSld^2 + \Slc^2  \x^2 u^8)(\sigma_\pm + \da)(\sigma_\pm + \overline \da),
\nonumber  
\end{align}
which is equation (\ref{eq:sigma_pm'}) in the text.

\subsection{$\mathbf{Sl(2,R)}$ transformation of conductivity \label{app:modular}}

Using (\ref{eq:Sl2R}) and (\ref{eq:Sl2onE}) in (\ref{eqa:sigma_pm})
we can determine how the conductivity transforms under $Sl(2,{\bm R})$.
In an orthonormal basis
\[\delta {\bm E}_i = \delta E_i + i \delta B_i \quad \longrightarrow \quad  (\Slc \da + \Sld) \delta {\bf E}_i\]
from which
\begin{align*}
\E_\pm + i \B_\pm &  \rightarrow \quad (\Slc \da + \Sld) (\E_\pm + i \B_\pm)\\
\E_\pm  - i \B_\pm &  \rightarrow \quad (\Slc \overline \da + \Sld) (\E_\pm - i \B_\pm)\\
\Rightarrow \qquad \qquad \E_\pm  &  \rightarrow \quad 
\left\{\Slc\left(\frac{\da + \overline\da}{2}\right) + \Sld\right \} \E_\pm 
+ i \Slc\left(\frac{\da - \overline \da}{2}\right)  \B_\pm\\
\B_\pm  &  \rightarrow \quad 
- i \Slc \left(\frac{\da - \overline\da}{2}\right)   \E_\pm
+ \left\{\Slc\left(\frac{\da + \overline\da}{2}\right) + \Sld \right\} \B_\pm.
\end{align*}
Hence
\[ \frac{\B_\pm}{\E_\pm} \quad \rightarrow \quad
\frac{\Slc\left( \frac{\da - \overline\da}{2 i} \right)  + 
  \left\{
   \Slc\left(\frac{\da + \overline\da}{2}\right) + \Sld \right\}\left(\frac{\B_\pm}{\E_\pm}\right) }
{ \left\{\Slc\left(\frac{\da + \overline\da}{2}\right) + \Sld \right\}
  - \Slc \left(\frac{\da - \overline\da}{2 i}\right)\left(\frac{\B_\pm}{\E_\pm}\right) }.
\]
With
\[\Sigma_\pm  = e^{-\lambda \phi}\frac{\B_\pm}{\E_\pm} \]
the $Sl(2,{\bm R})$ transformation (\ref{eq:Sl2R})
\[ e^{-\lambda \phi} = \frac{\da - \overline \da}{2 i} \quad
\longrightarrow \quad
\frac{1}{|\Slc\da + \Sld |^2}\left(\frac{\da - \overline \da}{2 i}\right)\]
leads to
\beq
\Sigma_\pm \quad  \longrightarrow
 \quad \frac{1}{|\Slc\da + \Sld |^2}
 \left(\frac{ \left\{\Slc\left(\frac{\da + \overline\da}{2}\right) + \Sld \right\} \Sigma_\pm  + \Slc \left(\frac{\da - \overline\da}{2}\right)^2 }
{\Slc\Sigma_\pm +
  \left\{\Slc\left(\frac{\da + \overline\da}{2}\right) + \Sld \right\} }\right).
\label{eq:Sigmatransform}
 \eeq
Now
\[ \Sigma_\pm = \sigma_\pm +\lambda \chi = \sigma_\pm + \left(\frac{\da + \overline \da}{2}\right)\]
and using this on the right hand side of (\ref{eq:Sigmatransform}) gives
\[ \Sigma_\pm  \longrightarrow
 \quad \frac{1}{|\Slc\da + \Sld |^2}
 \left(\frac{ \left\{\Slc\left(\frac{\da + \overline\da}{2}\right) + \Sld \right\} \sigma_\pm  + \Slc \left(\frac{\da^2 + \overline\da^2}{2}\right)
+\Sld\left( \frac{\da + \overline \da}{2} \right) }
{\Slc(\sigma_\pm + \da + \overline \da) + \Sld}\right).\]
Lastly
\[ \frac{\da + \overline \da}{2} \quad \rightarrow
\quad \frac{\Sla \Slc \da \overline \da + \frac{1}{2}(\Sla \Sld + \Slb\Slc )(\da + \overline \da) + \Slb\Sld }{|\Slc\da + \Sld|^2}\]
and, after some algebra,
\beq \sigma_\pm = \Sigma_\pm - \left( \frac{\da + \overline \da}{2}\right)
\quad \longrightarrow \quad
\frac{A \sigma_\pm + B}{C \sigma_\pm + D}\label{eq:sigmaSl2r}\eeq
with
\begin{align}
A & = \frac{\Slc (\da + \overline \da) + \Sld 
- \Sla |\Slc\da + \Sld|^2 + \Sld}{ |\Slc\da + \Sld|^2},\nonumber \\
B&= 
\frac{\Bigl\{\Slc (\da + \overline \da)+\Sld \Bigr\}
\Bigl\{ \Slc (\da + \overline \da) + \Sld - \Sla |\Slc\da + \Sld|^2 + \Sld\Bigr\}-|\Slc\da + \Sld|^2}{\Slc |\Slc\da + \Sld|^2},\nonumber\\
C&=\Slc ,\nonumber\\
D&=\Slc (\da + \overline \da) + \Sld.
\end{align}
Some further algebra shows that $A D - B C =1$,
this is in fact an $Sl(2,{\bf R})$ transformation (since $\da$ depends continuously on $u$ it is not $Sl(2,{\bf Z})$ for general $u$).

On the event horizon, with $\overline \da_1 = \overline \da_h$, $A_1 = A_{u=1}$, {\it etc}.,
and   $\sigma_+ = - \overline \da_1$,
\begin{align*}
 C_1 \sigma_+ + D_1 &= -C_1 \overline \da_1 +D_1=c\da_1 +d,\\
  A_1 \sigma_+ + B_1 &= -A_1 \overline \da_1 +B_1= \frac{(c\tau_1 + d)A_1 -1}{c}\\
  \Rightarrow \qquad \sigma_+ \quad \rightarrow \quad
  \frac{A_1\sigma_+ + B_1}{C_1 \sigma_+ +D_1} & =
  \frac{-A_1\overline \da_1 + B_1}{-C_1 \overline \da_1 +D_1}
  = -\left(\frac{a \overline \tau_1 + b}{c \overline \da_1 +d}\right)
\end{align*}
and similarly
\[ \sigma_- = - \da_1 \quad \rightarrow  \quad
-\left(\frac{\Sla \da_1 + \Slb }{\Slc \da_1 + \Sld}\right). \]

\section{Properties of Jacobi $\vartheta$-functions and modular forms\label{app:modular-forms}}

We collect together some useful properties of $\vartheta$-functions.
The definitions are those of \cite{WW} and most of the formulae
here are proven in that reference.  
The three Jacobi $\vartheta$-functions relevant to the analysis are defined as 
\begin{align}
\vartheta_2(\sigma)&=2\sum_{n=0}^\infty q^{(n+{\frac 1 2})^2}=
2q^{\frac 1  4}\prod_{n=1}^\infty\bigl(1-q^{2n}\bigr)\bigl(1+q^{2n}\bigr)^2,\\
\vartheta_3(\sigma)&=\sum_{n=-\infty}^\infty q^{n^2}=
\prod_{n=1}^\infty\bigl(1-q^{2n}\bigr)\bigl(1+q^{2n-1}\bigr)^2,\\
\vartheta_4(\sigma)&=\sum_{n=-\infty}^\infty (-1)^nq^{n^2}=
\prod_{n=1}^\infty\bigl(1-q^{2n}\bigr)\bigl(1-q^{2n-1}\bigr)^2, 
\end{align}
where $q:=e^{i\pi\sigma}$.

These three $\vartheta$-functions are not independent but are related by
\begin{equation}
\vartheta_3^4(\sigma)=\vartheta_2^4(\sigma) + \vartheta_4^4(\sigma).
\end{equation}
The following relations can be used to determine their
properties under modular transformations:
\begin{equation}
\vartheta_2(\sigma + 1 )=e^{i\pi/4}\vartheta_2(\sigma), \qquad
\vartheta_3(\sigma+1)=\vartheta_4(\sigma), \qquad
\vartheta_4(\sigma+1)=\vartheta_3(\sigma),
\label{eq:Tshift}
\end{equation}
\begin{eqnarray}
\vartheta_2(-1/\sigma)&=&\sqrt{-i\sigma}\;\vartheta_4(\sigma), \nonumber \\
\vartheta_3(-1/\sigma)&=&\sqrt{-i\sigma}\;\vartheta_3(\sigma), \label{eq:Sduality}\\
\vartheta_4(-1/\sigma)&=&\sqrt{-i\sigma}\;\vartheta_2(\sigma). \nonumber
\label{Stransform}
\end{eqnarray}
%
%
%
%
%

At the special points $\sigma=e^{i\pi/2}$ and $\sigma=e^{i\pi/3}$
the $\vartheta$-functions have the values
\begin{align}
\vartheta_3^2(e^{i\pi/2)})&=\sqrt{2}\vartheta_2^2(e^{i\pi/2)})=
\sqrt{2}\vartheta_4^2(e^{i\pi/2)})
=\frac{2}{\pi}K\left(\sin\left(\frac{\pi}{4}\right)\right), \label{eq:sigma_2}\\
e^{-i\pi/4}\vartheta_2^2(e^{i\pi/3})
&=e^{-i\pi/12}\vartheta_3^2(e^{i\pi/3})=
e^{i\pi/12}\vartheta_4^2(e^{i\pi/3})=\frac{2}{\pi}
K\left(\sin\left(\frac{\pi}{ 12}\right)\right), \label{eq:sigma_3}
\end{align}
where $K(k)$ is the complete elliptic of the second kind:
$K\bigl(\sin(\pi/4)\bigr)=\frac {1}  {4\sqrt{\pi}}(\Gamma(1/4))^2$, with
$\Gamma(1/4)\approx 3.6256$ the Euler $\Gamma$-function evaluated
at $1/4$, and $K\bigl(\sin(\pi/12)\bigr)\approx 1.5981$.

The $\vartheta$-functions have the following asymptotic forms 
\begin{eqnarray}
\sigma\rightarrow i\infty&&:\quad 
\vartheta_2(\sigma)\approx 2\;e^{\frac{i\pi\sigma}{4}} \rightarrow 0,
\quad \vartheta_3(\sigma)\rightarrow 1,\quad 
\vartheta_4(\sigma)\rightarrow 1;\\
  {\sigma} \rightarrow 0&&:\quad \vartheta_2(\sigma)\approx  \sqrt{\frac{i}{\sigma} },
\quad \vartheta_3(\sigma)\approx\sqrt{\frac{i}{\sigma}},\quad 
\vartheta_4(\sigma)\approx 2\sqrt{\frac{i}{\sigma}}\;e^{-{\frac{i\pi}{4\sigma}}}\rightarrow 0.\nonumber
\end{eqnarray}
In addition they satisfy the following
differential equations (see \cite{Rankin}, p.231, equation (7.2.17)),
\begin{eqnarray}
\frac{\vartheta_3^\prime}{\vartheta_3}-\frac{\vartheta_4^\prime}{\vartheta_4}
&=&\frac{i\pi}{4}\vartheta_2^4,\nonumber \\
\frac{\vartheta_2^\prime}{\vartheta_2}-\frac{\vartheta_3^\prime}{\vartheta_3}
&=&\frac{i\pi}{ 4}\vartheta_4^4,\label{eq:thetadot} \\
\frac{\vartheta_2^\prime}{\vartheta_2}-\frac{\vartheta_4^\prime}{\vartheta_4}
&=&\frac{i\pi}{ 4}\vartheta_3^4.\nonumber
\end{eqnarray}
Klein's $J$-invariant is defined as
\beq J = \frac{(\vtheta_2^8+\vtheta_3^8+\vtheta_4^8)^3}{54 \,\vtheta_2^8\vtheta_3^8\vtheta_4^8},\label{eq:J-def}\eeq
with the small $q$ expansion
\[ J = \frac{1}{12^3}\Bigl(q^{-2} + 744 + 196884 q^2 + 21493760\,{q}^{4} + O \bigr( {q}^{6} \bigr) \Bigr).\]
$J$ takes all complex values once and only once in the fundamental domain for $\Gamma(1)$.

Apart from $q=0$ there are two other fixed points of $\Gamma(1)$ in the fundamental domain, at
$\sigma=e^{i\pi/2}$  and $e^{i\pi/3}$, and near these $J$ has the expansions
\begin{align}
J(e^{i\pi/2} + \epsilon) & = 1 - \frac{3}{64 \pi^4 } \left\{\Gamma\left(\frac{1}{4}\right)\right\}^8 \epsilon^2 + \cdots,\\ 
J(e^{i\pi/3} + \epsilon) & = \frac{256}{(\sqrt{3} \pi)^3 } \left\{K\left(\sin\left(\frac{\pi}{12}\right)\right)\right\}^8 \epsilon^3 + \cdots
\end{align}
(the co-efficients can be calculated using (\ref{eq:sigma_2}), (\ref{eq:sigma_3}) and (\ref{eq:thetadot})).

The derivative of $J$ yields
\beq \frac{1}{2 \pi i J}\frac{d J}{d \sigma} =
\frac{(\vtheta_2^4+ \vtheta_3^4)(\vtheta_2^4 - \vtheta_4^4)(\vtheta_3^4+ \vtheta_4^4)}
{(\vtheta_2^8+\vtheta_3^8+\vtheta_4^8)},\label{eq:J'}\eeq
which transforms as ${J'}/{J}\rightarrow (c \sigma + d)^2 {J'}/{J}$,
it is the inverse of a modular from of weight $-2$.
Since $\Gamma(1)$ is generated by
\[\bm S = \bpm 0 & -1 \\ 1 & 0 \epm: \ \sigma \rightarrow -\frac{1}{\sigma}
  \qquad \mbox{and} \qquad 
  \bm T = \bpm 1 & 1\\ 1 & 0 \epm: \ \sigma \rightarrow \sigma + 1\]
the invariance of $J$ under $\Gamma(1)$ can be proven by using (\ref{eq:Tshift}) and (\ref{eq:Sduality})
while equation (\ref{eq:J'}) follows from (\ref{eq:thetadot}).

If
\[\gamma=\bpm \ia&\ib\\ \ic&\id \\ \epm \in \Gamma(1)\approx Sl(2,{\bf Z})/{\bm Z}_2\]
then the set of matrices
\[\Gamma(2)=\left\{\bpm  1&0\\  0 & 1  \epm   \mod\  2\right\}: \qquad
  (\ia, \id \ \mbox{both odd}; \ \ib, \ic \ \mbox{both even}) \] 
is a normal subgroup of $\Gamma(1)$.
A proper subgroup of $\Gamma(1)$ that contains $\Gamma(2)$ is called a level 2 subgroup and there are
four of these, apart from  $\Gamma(2)$ itself, three of which are referred to in the text,
\begin{align*}
\Gamma_0(2)= \left\{\begin{pmatrix} 1 & * \\ 0 & 1 \end{pmatrix} \mod 2\right\}:
\quad & (\ic \ \hbox{even}),\\
\Gamma^0(2)= \left\{\begin{pmatrix} 1 & 0 \\ * & 1 \end{pmatrix} \mod 2\right\}:
\quad & (\ib \ \hbox{even}).\\
\Gamma_\theta =\left\{\begin{pmatrix} 1 & 0 \\ 0 & 1 \end{pmatrix} \mod 2\right\} \cup
\left\{\begin{pmatrix} 0 & 1 \\ 1 & 0 \end{pmatrix} \mod 2\right\}:\quad  & (\ia\ic \ \mbox{even}).\\
\end{align*}   
where $*$ denotes either parity (the notation is that of \cite{Koblitz}).
A fifth level 2 subgroup, generated by $\bm{S T}$ and $\bm{T S}$, is not used. 

The subgroup $\Gamma_0(2)$ is generated by $\bm{S T}^2 \bm S$ and $\bm T$ and the function
\[ f = -\frac{\vtheta_3^4 \vtheta_4^4}{\vtheta_2^8}= -\frac{1}{256 q^2} \prod_{n=1}^\infty \frac{(1-q^{4n-2})^8}{(1+q^{2 n})^{16}}\]
is invariant under $\Gamma_0(2)$ and plays the same role for $\Gamma_0(2)$ as $J$ does for $\Gamma(1)$.
It has the small $q$ expansion
\[ f = -{\frac{1}{2^8}}\Bigl(q^{-2}-24+276\,{q}^{2} -2048 q^4+O \left( {q}^{6} \right) \Bigr).
\] 
As $\sigma$ runs from $0$ to $i\infty$ vertically up the imaginary axis $f$ decreases monotonically
from $0$ to $-\infty$.
Near $\sigma=0$ (an indeed $\sigma$ equal to any integer)
\[f(\epsilon) \approx -16 e^{-i\pi/\epsilon}. \]

The fundamental domain for $\Gamma_0(2)$ can be taken to be the vertical strip above the semi-circular
arc of radius $\frac 1 2$ spanning $\sigma=0$ and $\sigma=1$ and $f$ takes all complex values once
and only once in this fundamental domain.

There are fixed points of $\Gamma_0(2)$ at
at $\sigma_*=\frac{1+i}{2}$ and its images, where
\[ f( \sigma_* + \epsilon)= \frac{1}{4} - \frac{1}{64 \pi^4} \left\{\Gamma\left(\frac{1}{4}\right)\right\}^8 \epsilon^2 +\cdots. \] 

The invariant $f$  has the property that
\[ \frac{i}{\pi f}\frac{d f}{d \sigma} = \vtheta_3^4 + \vtheta_4^4\] 
transforms as ${f'}/{f}\rightarrow (c \sigma + d)^2 {f'}/{f}$,
it is the inverse of a modular form of weight $-2$ for $\Gamma_0(2)$.

\end{document}